\documentclass[manuscript,screen]{acmart}

\AtBeginDocument{%
  \providecommand\BibTeX{{%
    \normalfont B\kern-0.5em{\scshape i\kern-0.25em b}\kern-0.8em\TeX}}}

\setcopyright{acmcopyright}
\copyrightyear{2018}
\acmYear{2018}
\acmDOI{XXXXXXX.XXXXXXX}

\acmConference[Conference acronym 'XX]{Make sure to enter the correct
  conference title from your rights confirmation emai}{June 03--05,
  2018}{Woodstock, NY}
\acmPrice{15.00}
\acmISBN{978-1-4503-XXXX-X/18/06}

\usepackage{stfloats}
\usepackage{url}
\usepackage{caption}
\usepackage{subcaption}
\usepackage{tikz}
\usepackage{graphicx}
\usepackage{amsmath}
\usepackage{amsfonts}

\usepackage{amssymb}
\usepackage{mathtools}
\usepackage[ruled,vlined,linesnumbered]{algorithm2e}

\usepackage[utf8]{inputenc}
\renewcommand{\arraystretch}{0.4}

\begin{document}

\title{VL-DNA: Enhance DNA Storage Capacity with Variable Payload (Strand) Lengths}

\author{Yixun Wei}
\email{Yixun09@gmail.com}
\affiliation{%
  \institution{Department of Computer Science and Engineering, University of Minnesota}
   \country{USA}
}

\author{Wenlong Wang}
\affiliation{%
    \institution{Department of Computer Science and Engineering, University of Minnesota}
       \country{USA}
}

\author{Huibing Dong}
\affiliation{%
   \institution{Department of Computer Science and Engineering, University of Minnesota}
    \country{USA}
}

\author{Bingzhe Li }
\affiliation{%
 \institution{Department of Electrical and Computer Engineering, Oklahoma State University}
  \country{USA}}

\author{David H.C. Du}
\affiliation{%
  \institution{Department of Computer Science and Engineering, University of Minnesota}
   \country{USA}}

\renewcommand{\shortauthors}{Yixun Wei, et al.}

\begin{abstract}
DNA storage is a promising archival data storage solution to today’s big data problem. A DNA storage system encodes and stores digital data with synthetic DNA sequences and decodes DNA sequences back to digital data via sequencing. For efficient target data retrieving, existing Polymerase Chain Reaction (PCR) based DNA storage systems apply primers as specific identifier to tag different set of DNA strands. However, the PCR based DNA storage system suffers from primer-payload collisions, causing a significant reduction of storage capacity. This paper proposes using variable payload/strand length, which takes advantage of the inherent payload-cutting process, to split collisions and recover primers. The executing time of our scheme is linear to the number of primer-payload collisions. The scheme serves as a post-processing method to any DNA encoding scheme. The evaluation of three state-of-the-art encoding schemes shows that the scheme can recover thousands of usable primers and improve tube capacity ranging from 18.27\% to 19x.
\end{abstract}



\received{xx February xxxx}
\received[revised]{xx March xxxx}
\received[accepted]{xx June xxxx}

\maketitle

\section{Introduction}
Modern archival storage has been trying to preserve the ever-growing digital data reliably for centuries~\cite{miller2020future}. However, typical archival storage media/devices are not matching the booming storage demand. The storage requirement in hyper-scale data centres is expected to reach 32.6 million petabytes in 2030, which might surpass the total supplied storage capacity~\cite{alliance2021preserving}\cite{IDC}. Besides, the typical storage media/devices are also not durable enough (usually within a decade), thus introducing expensive data preservation costs.

DNA is emerging as a promising archival storage medium to meet the burgeoning storage demand. Theoretically, DNA storage can have a density of about 1 exabyte/mm$^3$  and preserve the data for hundreds of years~\cite{bornholt2016dna}. Even considering the implementation overheads of a practical archival system, existing DNA storage systems can still be orders of magnitude denser than tape\cite{alliance2021preserving}. Additionally, other unique properties of DNA, such as sustainability and energy efficiency, further make DNA a desirable medium for storing archival data for many decades.

In existing DNA storage systems, digital data is encoded and synthesized as the payloads of DNA strands. The DNA strands are usually dehydrated and stored in physical tubes for long-term preservation. When retrieving data, people liquidize the DNA and use a drop of the liquid for sequencing (i.e., the read process in DNA storage). To ensure efficient data retrieval, polymerase chain reaction (PCR) based random access has been introduced to DNA storage~\cite{organick2018random}\cite{bornholt2016dna}\cite{yazdi2015rewritable}\cite{yazdi2017portable}. In the PCR-based random access, each DNA payload is flanked by a pair of primers (i.e., short nucleic acid sequences) to form DNA strands. The primers serve as unique tags to identify different sets of DNA strands. When retrieving payloads (data), only strands with specific primer pairs can be amplified by PCR and then sequenced. In such a system, the number of DNA payloads stored in a tube is proportional to the number of usable primers in a tube. In this paper, the usable primers refer to primers that function well without violating bio-constraints, including no primer-payload collisions, which we discuss later.

The existence of primer-payload collisions disables many primers. Primer-payload collisions refer to almost identical subsequences between a primer and any portion of a payload~\cite{organick2018random}. A primer must be disabled for reliable PCR if it collides with any payloads in the same tube. The number of usable primers in a tube can decrease up to 70\%-99\% as the number of payloads increases~\cite{wei2022dna}. As a result, the tube capacity of random-access-based DNA storage is significantly decreased.

A few works are trying to increase the number of primers that can be used. For example,  the approach of nested primer~\cite{song2019multidimensional}\cite{yamamoto2008large} combines multiple short usable primers to form a new, longer primer. The number of new primers can increase by combining the multiple layers of short usable primers. However, the nested primer approach does not solve the primer-payload collision issue directly. Besides, using the usable primers in a nested manner will reduce each DNA strand's payload length, assuming the length is fixed. Other related work~\cite{organick2018random}\cite{el2022high} keep randomizing DNA sequences until all payload sequences have no collision with primers and still satisfy biological constraints. These works can avoid collisions in small-scale data. Nevertheless, as data storage scales, they have an unacceptable computation overhead and a giant mapping table to record the randomization.

This paper proposes a variable payload (strand) length scheme, called VL-DNA, to break up primer-payload collisions using DNA storage's inherent payload-cutting process. Instead of cutting DNA payloads with fixed current maximum length (e.g., 200 bases), VL-DNA cuts payloads into several variable payload lengths (e.g., 150, 160, 190, and 200 in this paper). We combine these payload lengths to build cutting points at different positions so that we can break the potential primer-payload collisions and recover more usable primers. Since strand length is the payload length plus metadata length (including primers and an internal index), variable payload length means variable strand length. Hereafter, we use variable payload/strand length interchangeably. A heuristic scheme is proposed to find the cutting points of payload to balance the trade-off between DNA strand length and the number of recovered usable primes. As a result, VL-DNA extricates a considerable amount of primers from collisions and thus improves the tube capacity of DNA storage. Note that VL-DNA is generic and independent. We can apply it to any existing DNA encoding schemes or optimizations (e.g., nested primer approach) to gain storage capacity enhancements.

The rest of this paper is organized as follows. Section~\ref{sec:backgroud} introduces DNA storage backgrounds, including storage workflow, factors that affect capacity, and primer-payload collision. Section~\ref{sec:3} elaborates on the concept of the variable payload length and the corresponding algorithms. Section~\ref{sec:complete scheme} shows the complete DNA storage system equipped with the VL-DNA scheme. Section~\ref{sec:evaluation} evaluates our scheme. Conclusions are drawn in Section~\ref{sec:conclusion}.

\section{Background}

\label{sec:backgroud}
\subsection{DNA Storage in Brief}

\begin{figure*}[htbp]
\centering
\small
\includegraphics[width=\textwidth]{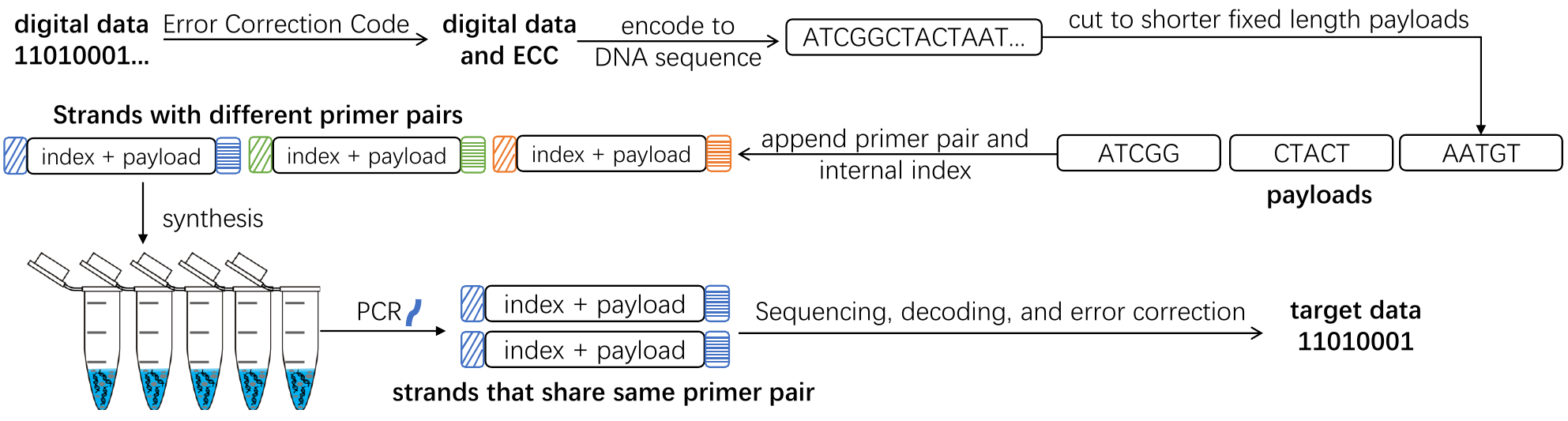}
\caption{Workflow of typical DNA storage system}
\label{fig:DNA storage workflow}
\end{figure*}

Several researchers have demonstrated the feasibility and discussed the potential capacity of DNA archival storage~\cite{DNAfuture}~\cite{li2020can}\cite{wei2022dna}. Figure~\ref{fig:DNA storage workflow} shows the typical workflow of the current DNA storage system.

In existing DNA storage systems, digital data will first be added with some error correction code (e.g., Reed–Solomon Code) for data recoverability~\cite{lin2022managing}~\cite{carmean2018dna}. Certain encoding schemes will then encode the resultant data to a sequence of DNA bases (A, T, G, and C). The DNA sequences will be cut into multiple shorter subsequences called DNA payloads. Each payload will be flanked by a pair of primers and an internal index to form a DNA strand. The primer pair is a unique tag that enables random access to DNA storage based on PCR. Multiple DNA strands can share one primer pair, and an internal index helps identify the strands. Eventually, each DNA strand will be chemically synthesized base by base and stored in physical tubes.

\begin{figure}[htbp]
  \begin{subfigure}[b]{0.45\textwidth}
    \includegraphics[width=\textwidth]{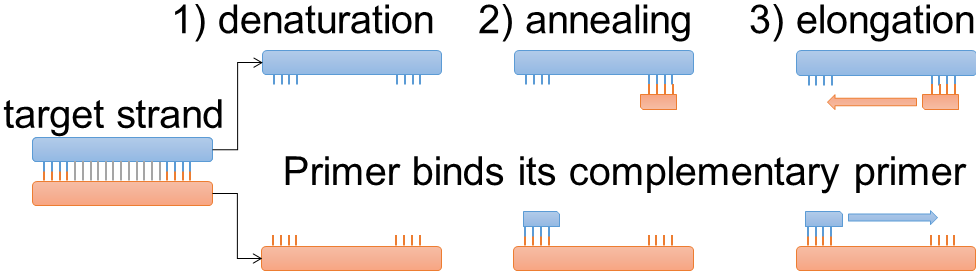}
    \caption{standard PCR}
    \label{fig:standard PCR}
  \end{subfigure}
  \hfill
  \begin{subfigure}[b]{0.45\textwidth}
    \includegraphics[width=\textwidth]{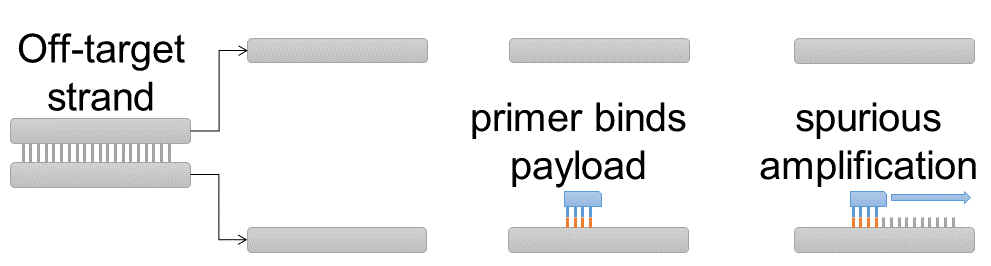}
    \caption{PCR with primer-payload collision} 
    \label{fig:primer-payload collision}
  \end{subfigure}
  \caption{standard PCR and defective PCR with primer-payload collision}
\end{figure}

The PCR-based random access is a process that first amplifies target DNA strands with a specific primer pair and then sequences only the target strands. The amplification usually takes 20 to 40 PCR cycles. A standard PCR cycle is shown in Figure 2 (a). In each cycle, target DNA strands are denatured into two single-stranded DNA molecules, and each molecule anneals with the added complementary primer templates. Then the DNA bases (A, T, C, G) from the added mixture solution will pair with the annealed molecules to form new DNA strands. After multiple PCR cycles, the target strands have enough concentration and can be sequenced out. Finally, the read-out DNA sequences can be decoded back to digital data. The ECC added before can help recover some errors during the synthesis and sequencing processes.

\subsection{DNA Tube Capacity and Primer-Payload Collisions}
\label{sec:tube capacity}
A DNA storage system must obey many bio-constraints to provide reliable data storage; thus, many factors affect DNA tube capacity. DNA tube capacity is the multiplication of the following factors~\cite{li2020can}\cite{wei2022dna}: 1) DNA payload length, 2) the number of digital bits each DNA base can store (i.e., encoding density), 3) the number of strands that one primer pair can accommodate (i.e., parallel factor), and 4) the number of usable primers. A practical DNA strand length is limited to 100-300 bases because the error rate soars up when the strand becomes overlong~\cite{alliance2021preserving}\cite{matange2021dna}. After excluding the primer pair and internal index, a payload is usually no longer than 200 bases. Similarly, the parallel factor is also limited (e.g., 1.55 $\times 10^6$~\cite{li2020can}). Based on different encoding schemes, the encoding density varies from less than one bit per DNA base to at most two bits per DNA base (e.g., A=00, T=01, C=10, G=11). The number of usable primers is also limited by certain primer design rules and the primer-payload collisions that we will discuss later. The DNA tube capacity equals the payload length $\times$ encoding density $\times$ parallel factor $\times$ the number of usable primers / 2 (i.e., the number of primer pairs).

\begin{figure}[!t]
\centering
\includegraphics[width=0.48\textwidth]{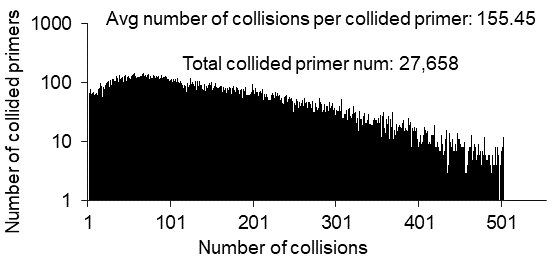}
\caption{Distribution of primers with different numbers of collisions (encoding scheme: Blawat, data: 135MB video)}
\label{fig:collision distribution}
\end{figure}

Given all biological constraints, primer-payload collision is the most critical factor affecting the tube capacity. As Figure 2 (b) shows, a collision occurs if a pair of almost identical subsequences exists between a primer and a portion of payload stored in the same tube~\cite{organick2018random}\cite{wei2022dna}. The subsequences are usually longer than 12 bases and have at most two mismatches or gaps. In each PCR cycle, the PCR can amplify some irrelevant strands if these strands contain payloads that collide with the target primer. More importantly, this consumes the limited PCR reagents (e.g., complementary primers used to bind with the target primer and free DNA bases waiting for complementing new double-helix strands). Figure~\ref{fig:collision distribution} shows the distribution of primers with different numbers of payload collisions. The input data is a 135MB video~\cite{135MBVideo}, and the encoding scheme is Blawat code~\cite{blawat2016forward} (see the encoding scheme, primer generating, and collision checking in Section~\ref{sec:evaluation}). Although input data is only 135MB, 27,658 primers (i.e., 98.77\% of primers in our primer library) have collisions with those payloads, and the average number of collisions per collided primer is 155.45 (i.e., on average, a primer has collisions with 155.45 payloads). If the input data keeps scaling up, most collided primers likely have hundreds or thousands of collisions. In this case, PCR amplification is very vulnerable to primer-payload collisions. The target primer must compete for the limited PCR reagents with thousands of collided payloads in each PCR cycle. Even if a primer has only a few collisions at the beginning, the collisions can still significantly impact the final result. That is because PCR is an exponential augmentation process. Slight amplification variations can exacerbate cycle by cycle and consume ever more PCR reagents. Eventually, the resultant solution will be saturated with plenty of irrelevant sequences but very few target sequences. As a result, the sequencing for target sequences is inhibited.

Organick et al. first ~\cite{organick2018random} reported a PCR failure due to primer-payload collisions. They successfully accessed a target file when the DNA tube contained no other data but failed to do so when there were nine files in the tube (the target data is 17.4\% of the whole data in the tube). After that, Wei et al.~\cite{wei2022dna} investigated the impact of primer payload collisions based on 1.5TB of different types of digital data. Their results showed that the primer payload collision could cause 70\% to 99\% tube capacity loss depending on different encoding schemes. Most of the existing encoding schemes can achieve only several Gigabytes of tube capacity, which is much lower than people's expectations and may inhibit DNA storage from practical use.

Given that other capacity factors (e.g., strand length) are all restricted by current biotechnologies, it is critical to relieve the malevolent influence of primer-payload collisions to enhance DNA storage capacity. In this paper, we apply BLAST~\cite{BLAST} to check the primer-payload collisions. BLAST is one of the most widely used genomic sequence analysis tools and has been applied by previous DNA storage works~\cite{organick2018random}\cite{wei2022dna}.

\section{VL-DNA Design and Algorithm}
\label{sec:3}

In this section, we first explain how VL-DNA uses the inherent payload-cutting process to break up collisions. Then, we propose a heuristic algorithm to achieve the balance between payload length and primer recovery.	

\subsection{Variable Payload Length}
\label{sec: variable length}

\begin{figure}[htbp]
\centering
\small
\includegraphics[width=0.9\textwidth]{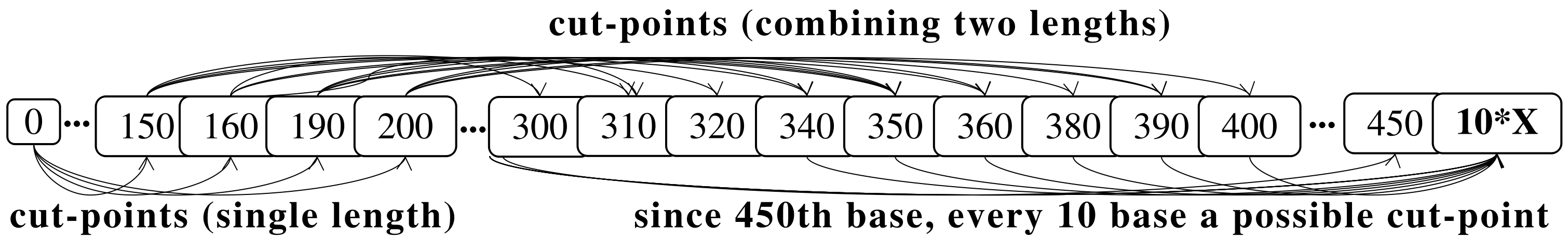}
\caption{Possible cut-points with payload length 150/160/190/200}
\label{fig:payload length}
\end{figure}

A collision occurs if there exist two consecutive and almost identical subsequences longer than 12 bases in a primer and a payload. We can remove a collision by splitting it into two parts no longer than 12 bases. However, current DNA storage systems cut DNA sequences into fixed-length payloads. In this case, we can only cut collisions located in the multiple of the fixed length. By contrast, arbitrary length can cut any collision. However, arbitrary length can generate very short payloads that not only waste strand length but are easily mistaken as noisy reads (i.e., by-products of sequencing, usually short broken DNA strands). Moreover, arbitrary length is less tolerable to DNA base insertion and deletion errors.

Making a trade-off between fixed and arbitrary payload lengths, we identified and proposed to use only four payload lengths as bases and combine them into longer variable lengths to cut most collisions. Given that the current practical strand length is usually 100-300 bases, we assume the maximum payload length is around 200 bases after excluding the primer pair and internal index. We select 150/160/190/200 as our four basic payload lengths. Fig.~\ref{fig:payload length} shows possible combinations of the four payload lengths. Starting from any position, we can set cut points at the next 150/160/190/200th base, generate one payload with a corresponding length, and break up collisions at corresponding positions. Combining two or more payloads, we can further break up collisions at more places. Specifically, we can have a possible cut point every ten bases since the 450th base. Since collisions are longer than 12 bases, a possible cut point of every ten bases can cover most of the collisions. We only need one DNA base as metadata to indicate the length of the payload (i.e., less overhead). 

\vspace{0.1cm}

\noindent\textbf{Why these four payload lengths: } except the payload length group of 150/160/190/200, there are also many other lengths. For example, if we have a payload length as short as 5, we can have cut points every five bases. If we adopt more than four payload lengths, we may have more length combinations and cover more collisions. However, too short or too many payload lengths can also bring other overheads. To comprehensively understand and select payload lengths, we summarize the following principles:
\begin{itemize}
    \item \textbf{Payload length should not be too short:} An abridged payload length means a loss of capacity. Meanwhile, DNA sequencing will generate plenty of noisy reads (i.e., broken segments of DNA strands). Short payloads could be easily mistaken as noisy reads. Therefore, we only select lengths in the range of [150,200] based on the assumption that the current maximum payload length is 200. 
    
    \item \textbf{The number of payload lengths should not be too many:} DNA strand needs extra metadata to indicate strand length since strand length is essential in error detection. In Section~\ref{sec:evaluation} we will see that increasing the number of payload lengths does not significantly increase the capacity and will bring higher length indicating overhead. Therefore,  this paper aims for four payload lengths with only one DNA base as the length indicator.   
    
    \item \textbf{The combination of payload lengths should cover as many areas as possible:} The more areas a payload length group can cover, the more chances collisions are cut. To estimate how many areas a payload length group can cover, we define \textit{number of covered bases} of a payload length group as the total number of bases if both two sides of the base can have possible cut points. The cut points on both two sides should be within 12 bases. Because collisions are longer than 12 bases, a base with potential cut points on both sides means collisions starting from this base can always be covered. 
\end{itemize}

Among all the four-length groups in the range of [150,200], 150/160/190/200 is one of the groups with the highest \textit{number of covered bases}. We compare the length group of 150/160/190/200 with other length groups in Section~\ref{sec:evaluation}. The result shows that our length group of 150/160/190/200 outperforms all other length groups with four lengths and is comparable to other length groups with more lengths, as shown in section~\ref{sec:evaluation}. 
 
Because the four basic lengths are multiple of 10, the combination is always a multiple of 10. We check positions inside each collision that are multiple of 10. The position is the collision's functional cut point if it splits the collision into two parts that are both shorter or equal to 12 bases. Otherwise, we use BLAST~\cite{BLAST} to double-check whether the collision is removed. We consider the position as the collision's functional cut point only if BLAST returns no collision. Otherwise, we consider that our variable length cannot remove the collision.

\subsection{Problem Formulation:}

We first define two properties of primers and then use the properties to formulate the primer selection problem.
\begin{itemize}	
    \item \textit{Primer\_capacity} is the number of DNA bases the primer can accommodate. Given a fixed parallel factor, \textit{Primer\_capacity} is determined by the payload length of DNA strands. Since VL-DNA cut payloads into variable (shorter) lengths, the more collisions a primer has, the smaller the \textit{Primer\_capacity} is.	
    	
    \item \textit{Primer\_conflicts} describes how many other primers a primer conflicts with. We consider two primers conflicting with each other when their collisions are closely grouped and cannot be all cut. Recovering a primer inhibits the recovery of other conflict primers.	
\end{itemize}	
\noindent\textbf{Problem definition:} Given all collided primers $V=(p_1,p_2,....p_{|V|})$, find a primer subset $P \subseteq V$ that $\sum_{p_i \in P} \textit{ Primer\_capacity }(p_i)$ is maximum, and for each $p_i \in P$, $\textit{Primer\_conflicts }(p_i) \cap P = \emptyset$. 

Consider an undirected graph in which each vertex is a collided primer, vertex weight is the \textit{Primer\_capacity}, and edges connect vertices if the corresponding primers conflict. We can interpret the problem as selecting a vertex subset in which the selected vertices have maximum total weight and have no edges connected. When the weight of all vertices is equal, this is the maximum independent set problem~\cite{xiao2017exact}, a well-known NP-hard problem. Since the maximum independent set problem is a particular case of our problem, any solution to our problem also works to the maximum independent set problem. Since exactly solving our problem is at least as hard as solving the maximum independent set problem, we consider solving our problem with efficient heuristic algorithms.

\subsection{VL-DNA Algorithm}
\label{sec: VL-DNA algorithm}
The heuristic algorithm is based on the number of collisions and conflicts each primer has. The number of collisions hints at the direct overhead of recovering a primer. The more collisions a primer has, the shorter payloads will generate when recovering the primer. The number of conflicts indicates the indirect overhead of recovering a primer. One primer's recovery will block its conflicted primers' recovery and lead to a potential overall capacity decrease.

\begin{algorithm}[htbp]
\SetAlgoLined
\SetKwInOut{KwInput}{Input}
\SetKwInOut{KwOutput}{Output}
\SetKwInOut{Parameter}{Parameter}
\SetKwFunction{Fvariablelen}{\textbf{Variable Length}}
\SetKwIF{If}{ElseIf}{Else}{if}{}{else if}{else}{}
\KwInput{Collided primers and associated collisions}
\KwOutput{Cut points}
\BlankLine

\SetKwProg{Fn}{Function}{:}{}
\Fn{\Fvariablelen{}}{
    
    recovered[]= Null;\ 
    
    abandoned[] = Null;\ 
    
    pending[] = all collided primers;\ 
    
    cut\_point[] = Null;\
    
    \BlankLine
    ascending sort primers in pending[];\ 
        
    
    \ForEach{primer $p_i$ in pending[]}{
        \BlankLine 
        \ForEach{$p_i$'s collision $c_i$}{
            cut\_point[].add($c_i$'s cut-point);\
        }
        
        \ForEach{$p_j$ that conflicts $p_i$}{
           move $p_j$ from  pending[] to abandoned[];\
        }
        move $p_i$ from pending[] to recovered[];\
    }
    \KwRet cut\_point[];
}
\caption{VL-DNA algorithm}
\label{alg:payload cutting}
\end{algorithm}

The algorithm is shown in Algorithm~\ref{alg:payload cutting}. The input is all collided primers and their associated collisions. The goal is to obtain the necessary cut points to remove collisions and recover primers, so that tube capacity is enhanced to the most extent. We initialize four lists: $recovered[]$, $pending[]$, $abandoned[]$, and $cut\_point[]$. All collided primers are initially in $pending[]$. We first sort primers in ascending order based on either their number of collisions or conflicts. After that, we sequentially try to recover each primer in $pending[]$: the functional cut point of each collision of the current primer $c_i$ will be added into $cut\_point[]$; the conflicted primers of the current primer will also be moved from $pending[]$ to $abandoned[]$. We then move the current primer from $pending[]$ to $recovered[]$.

After processing all primers, we return the $cut\_point[]$. We then combine the variable payload lengths to reach each cut point and cut payloads accordingly. Note that this algorithm only returns necessary cut points used to remove collisions. As for other places with no collisions, we can just cut payloads with a maximum allowed length or other lengths if required. The sorting criteria can be either collision number or conflict number because they have a very strong correlation (Section~\ref{sec:evaluation}).

\section{DNA storage system equipped with VL-DNA}
\label{sec:complete scheme}

\begin{figure}[htbp]
\centering
\includegraphics[width=0.6\textwidth]
{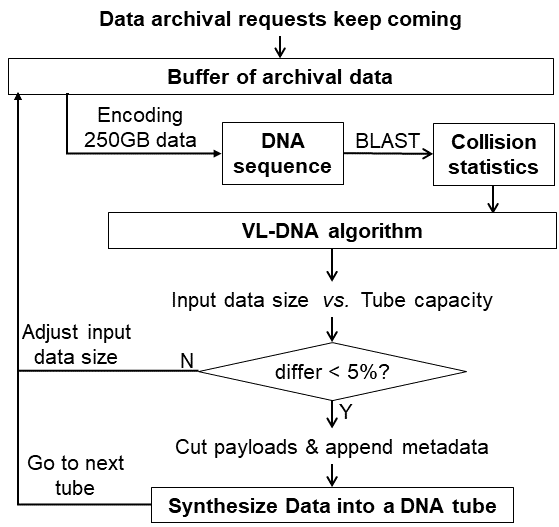}
\caption{Workflow of the DNA archival storage system equipped with VL-DNA}
\label{fig:workflow}
\end{figure}

An archival storage system (e.g., Amazon S3 Glacier) keeps receiving data archival requests. A DNA archival storage system equipped with the VL-DNA scheme should be able to ingest the sequential archival requests and store data tube by tube. We propose a system workflow to handle this situation, as shown in Fig.~\ref{fig:workflow}.

The incoming data would first be buffered until it is enough to fill a DNA tube. Initially, we start storing data in a DNA tube when it reaches 250 GB (i.e., an empirical value of DNA tube capacity). The storage system appends Error correction codes for the data and then encodes the data into a long DNA sequence. Next, the system uses BLAST to check collisions and obtain each collided primer's collisions and their conflicted primers. Based on the collision statistics, the VL-DNA algorithm intelligently suggests certain primers to recover and returns necessary cut points used to break up collisions of the recovered primers. The system can then calculate the number of payloads the cutting process will generate.

Meanwhile, the system can also calculate the number of storable payloads based on current input data. It is the multiplication of the number of usable primers, 0.5 (we need a primer pair), and parallel factor~\cite{li2020can}. The number of payloads generated and storable can indicate the difference between input data size and calculated tube capacity. Only when the storable payloads are within 5\% higher/lower than the generated payloads do we consider the VL-DNA algorithm's result reasonable. Otherwise, we adjust the input data size and rerun the algorithm. The size adjustment follows a binary search: each time, we adjust the input data size to the middle of the last input data size and the last storable data size. The rerunning is quite efficient because most digital data has already been encoded and collision-checked. We only need to rerun the algorithm itself. The algorithm's time complexity is O(n), where n is the number of collisions being cut.

After finding a proper input data size, the system cuts the DNA sequence into payloads based on the cut points defined by the VL-DNA algorithm. Finally, the system appends metadata (i.e., length indicator, primer pair, and internal index) for each payload and sends them to DNA synthesis. If there is still enough data in the buffer, the system prepares to process the data for the next DNA tube. Note that data is sequentially processed: we do not intentionally choose data and allocate them into different tubes. The VL-DNA scheme works independently with DNA encoding. Even if payloads are cut into different lengths, the decoding process is still the same: orderly assemble payloads based on the internal index and then use the corresponding encoding scheme to decode the DNA sequence back to a bit sequence. We can apply the VL-DNA scheme as a post-processing method to any encoding scheme to enhance the storage capacity.

\section{Evaluation}
\label{sec:evaluation}
This section evaluates the VL-DNA scheme in aspects of capacity improvements and payload length selection. Following previous works~\cite{organick2018random}\cite{wei2022dna}, we generate a primer library with 28,000 primers for the evaluation. To obtain DNA tube capacity, we adopt the 1.55 $\times 10^6$ as the parallel factor and apply the same capacity calculation as described in Section~\ref{sec:tube capacity}.

\subsection{Tube Capacity Improvements}
\label{sec:capacity improvement}
In this subsection, we use multiple encoding schemes to encode digital data and apply two VL-DNA schemes to evaluate the improvements in the number of usable primers and tube capacity. At the time this paper is completed, the only scheme that tries to alleviate the primer-payload collision problem is payload randomization used by Organick et al.~\cite{organick2018random}. The randomization is done by decoding collided payloads back to bit sequences and XOR the bit sequences with a number of pseudo-random bit sequences. The scheme keeps randomizing the input data until the randomized data will be encoded into a DNA sequence without any collision. This scheme works well only in small-scale data with a small-scale primer library. Given the size of our primer library (i.e., 28,000 primers), each payload has to be randomized many times to have no collision with any primer in the library. As the input data set keeps scaling, the randomization scheme consumes unacceptable time for collision checking with BLAST and DNA encoding/decoding. Randomization also needs a humongous randomization seed mapping table to record which randomization has been applied on a payload. 

To fairly evaluate the VL-DNA scheme, we designed three baselines and compare them with two VL-DNA schemes:
\begin{itemize}	
    \item \textbf{Fixed 200 bases:} Cut the encoded DNA sequence into multiple fixed 200 bases payloads
    	
    \item \textbf{Fixed 200 bases + 5 randomization opportunities:} We also cut the encoded DNA sequence into multiple fixed 200 bases payloads. After that, each payload can be randomized at most 5 times. We check the collisions of each randomization and select the randomized result that collides with the least number of primers. This is different from the original randomization scheme which allows randomizing a payload unlimited times.\footnote{Grass is encoded and decoded three DNA bases at a time which is not divisible by 200 bases. When decoding payloads generated by Grass code, we leave the last two bases unchanged and only randomize the first 198 bases.}

    \item \textbf{Fixed 200 bases + 10 randomization opportunities:} Same as the last baseline except that each 200 bases payload has 10 randomization opportunities.

    \item \textbf{VL-DNA (collisions):} Apply the VL-DNA scheme to cut payloads based on the collision statistics reported by BLAST~\cite{BLAST}. The procedure is shown in Figure.~\ref{fig:workflow}. The VL-DNA algorithm is collisions-based (i.e., prioritizes the recovery of primers with fewer collisions) as discussed in Section.~\ref{sec: VL-DNA algorithm}.	

    \item \textbf{VL-DNA (conflicts):} Same as VL-DNA (collisions) except that the VL-DNA algorithm is primer conflicts based (i.e., prioritizes the recovery of primers that have fewer conflicts with other primers).	 	
\end{itemize}

We have collected digital data from \textit{ImageNet}~\cite{imagenet_cvpr09}, \textit{LibriSpeech}~\cite{panayotov2015librispeech} and other small collections from \textit{InternetArchive}~\cite{InternetArchive}. The types of digital data have insignificant influence on collisions because all possible bit sequences will appear multiple times as data set scales~\cite{wei2022dna}. Therefore, we mixed the above data to mimic a common archival storage input. First, we apply Reed–Solomon(255,239) on the input digital data to append 16 parity bytes to each 239 data bytes. We then implemented three state-of-the-art encoding schemes: Rotation code~\cite{bornholt2016dna}\cite{yazdi2015rewritable}, Blawat code~\cite{blawat2016forward}, and Grass code~\cite{grass2015robust}, to encode the data respectively. 

\begin{figure}
\centering
\begin{subfigure}[b]{0.45\textwidth}
   \includegraphics[width=\textwidth,height=5cm]{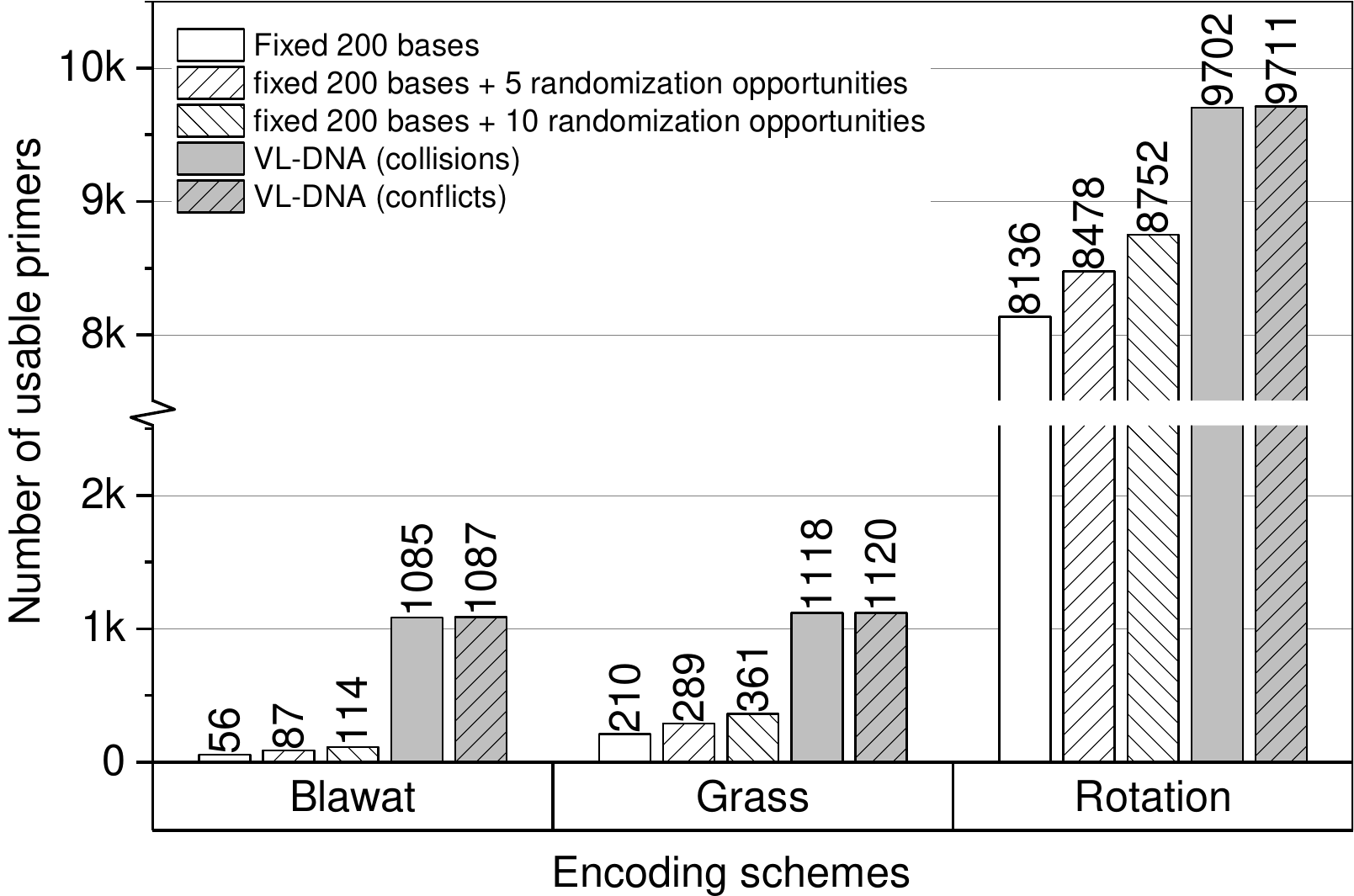}
   \caption{Number of usable primers}
   \label{fig:primer_enhancement} 
\end{subfigure}
\hfill
\begin{subfigure}[b]{0.45\textwidth}
   \includegraphics[width=\textwidth,height=5cm]{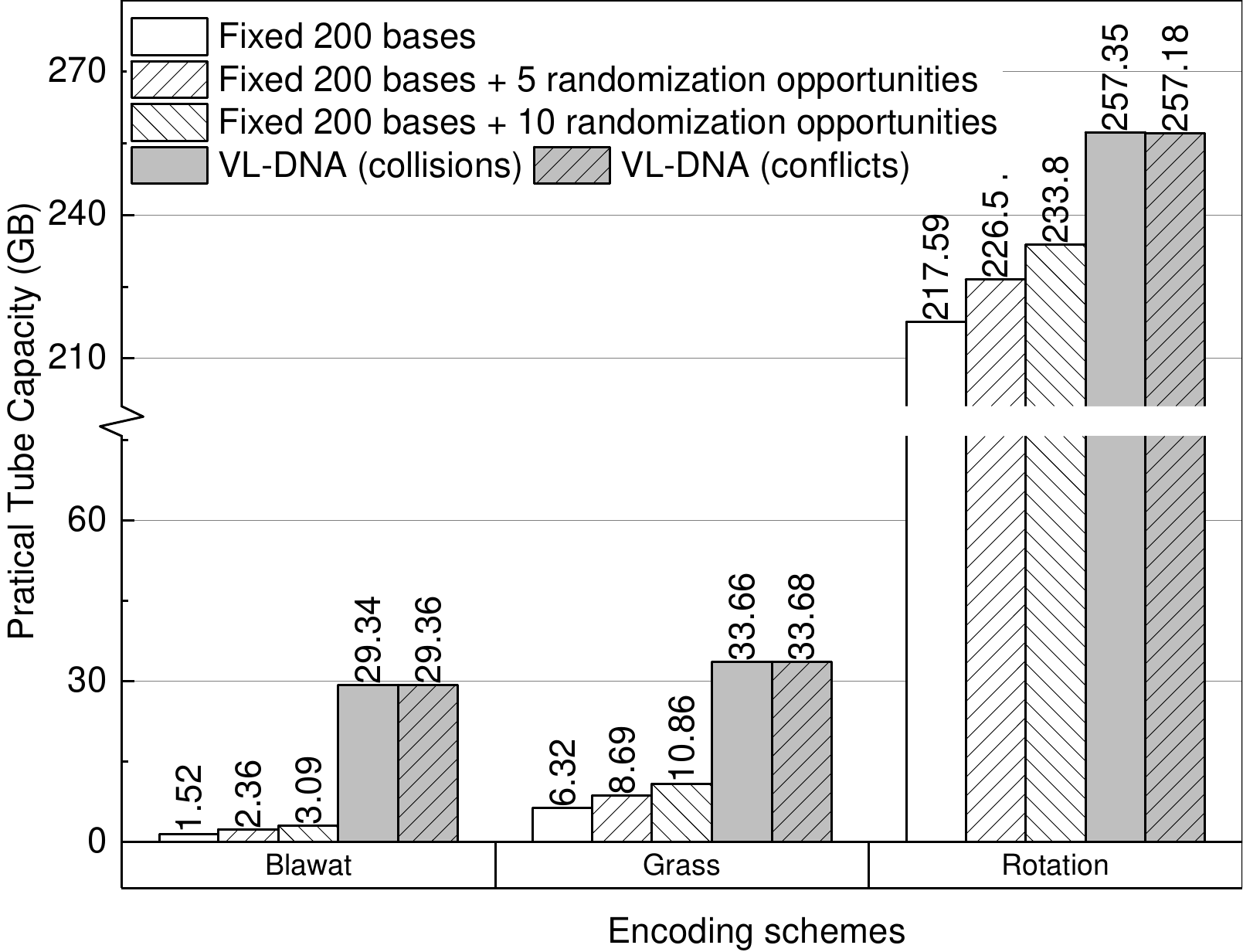}
   \caption{Achieved tube capacity}
   \label{fig:capacity_enhancement}
\end{subfigure}
\caption{Enhancements of VL-DNA in the number of usable primers and tube capacity. The parallel factor is 1.55 $\times 10^6$. The encoding density of Blawat/Grass/Rotation is 1.6, 1.78, 1.58.}
\end{figure}

We respectively perform the three baselines and two VL-DNA schemes on the encoded DNA sequences. The results are shown in Figure~\ref{fig:primer_enhancement}. Due to the difference in encoding schemes, Rotation code initially has more usable primers than Blawat and Grass (Rotation: 8,136, Blawat: 56, Grass: 210) when we only cut fixed 200 bases. That is because Rotation code creates a particular sequence pattern on its payloads: every DNA base will differ from the last base. This pattern makes the Rotation code avoid collisions with primers that have several identical consecutive bases. In comparison, Blawat code and Grass code do not restrict a string pattern on their payloads and thus suffer more collisions (see more analysis in ~\cite{wei2022dna}). This difference makes DNA payloads of Blawat code and Grass code have more collisions than DNA payloads of Rotation code. As a consequence, VL-DNA can recover more primers in Rotation code than Blawat code and Grass Code. Regardless of the different collision resistance of different encoding schemes, the VL-DNA scheme can still recover thousands of primers (e.g., Blawat: 1,029, Grass: 908, Rotation: 1,566 in the collisions-based VL-DNA). To recover those primers, the VL-DNA scheme (collisions-based) respectively cuts 0.94 million, 1.07 million, and 49.69 million collisions in payloads of Blawat (157 billion DNA bases), Grass (162 billion DNA bases), and Rotation (1,398 billion DNA bases). This indicates a rather acceptable overhead of the VL-DNA scheme. The average payload length of collisions-based VL-DNA is 199.87 for Blawat code, 199.85 for Grass code, and 198.54 for Rotation code. Although some payloads are tens of bases shorter than the maximum payload length (e.g., 160 vs 200), the VL-DNA algorithm only recovers primers with fewer collisions, and the large data set can amortize the length reduction of a few payloads. 

\begin{figure}[htbp]
\centering
\includegraphics[width=0.5\textwidth]{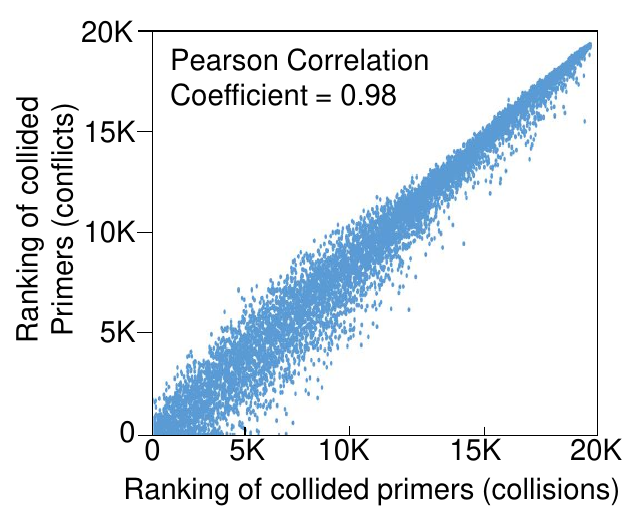}
\caption{Rank primers based on collision number and conflict primer number. The encoding scheme used is Rotation and the input data is \textit{ImageNet}.}
\label{fig:correlation}
\end{figure}

The recovery of usable primers enhances tens of Gigabytes of tube capacity as shown in Figure~\ref{fig:capacity_enhancement}. Compared with only fixed 200 bases, Blawat code and Grass code become much more practical after the VL-DNA scheme improves the tube capacity (i.e., 19x capacity increase for Blawat and 5x capacity increase for Grass). Rotation code also improves 18.27\% of its tube capacity despite the relatively large initial tube capacity. On the other hand, the two randomization schemes do not significantly increase the number of usable primers and tube capacity. That is because the randomization schemes are essentially equivalent to selecting the most collision-resistant bit sequence from 5 or 10 randomized bit sequences. Even though the selected randomized bit sequence after encoding can have fewer collisions than other randomized bit sequences, the selected bit sequence is still constrained by the encoding schemes. Due to the absence of collision resistance, Blawat code and Grass code still suffer from the collision problem even applied with randomization schemes. Rotation code, because of the collision resistance stemming from its intrinsic special payload pattern, benefits a little from the randomization schemes. However, the increase is still subtle compared with the VL-DNA scheme. If the number of primers in a primer library keeps increasing, the randomization scheme can provide less and less benefit.

When comparing the VL-DNA schemes based on either primer collisions or primer conflicts, the conflict-based scheme usually recovers a few more primers but has a shorter average payload length. However, the differences are negligible. That is because the primer collision number and primer conflict number have a very strong correlation, as shown in Figure~\ref{fig:correlation}. In Fig.\ref{fig:correlation}, X-axis is the ranking of collided primers based on the number of collisions, and Y-axis is the ranking based on the number of conflicts. The encoding scheme used is Rotation code~\cite{bornholt2016dna}, and the digital data is \textit{ImageNet}~\cite{imagenet_cvpr09}. The Pearson correlation coefficient, a measure of the linear dependence between the number of conflict primers and collisions, is 0.98. That means primers with many collisions also tend to have many conflicted primers. It is because conflicts come from the inappropriate position of collisions, and a primer with fewer collisions usually has a more negligible conflict probability. This observation allows us to prioritize primers either based on the number of collisions or the number of conflicted primers rather than any complicated combination schemes.

\subsection{Comparison of Payload Length Groups}
\label{sec:length group}
\begin{table}[htb]
\small
  \begin{center}
    \caption{Comparison of Length groups in terms of the number of usable primers and tube capacity}
    \label{tab:length group comparison}
     \tabcolsep=0.08in
     \renewcommand{\arraystretch}{1.3}
    \begin{tabular}{c|c|c} 
      \textbf{Length groups} & \textbf{ \# usable primers} & \textbf{Tube capacity} \\\hline
        \textbf{150/160/190/200} &  \textbf{9,702} &  \textbf{257.35 GB}\\ \hline
         -------------- & ---   & ----\\ \hline
         150/160/180/200 & 9,657 & 255.11 GB \\ \hline
         150/170/190/200 &  9,652 & 255.69 GB \\ \hline
         150/170/180/200 &  9,652 & 255.48 GB \\ \hline
         150/180/190/200 &  9,643 & 255.49 GB \\ \hline
         160/170/180/200 & 9,583 & 254.66 GB \\ \hline
         160/170/190/200 & 9,612 & 255.50 GB \\ \hline
         160/180/190/200 & 9,585 & 254.83 GB \\ \hline
         170/180/190/200 & 9,532 & 253.59 GB \\ \hline
         --------------- & ---   & ----\\ \hline
         160/170/180/190/200 &  9,628 & 255.80 GB \\ \hline
         150/160/170/180/190/200 & 9,756 & 258.96 GB \\ \hline
         --------------- & ---   & ----\\ \hline
         100/130/160/200 & 9,885 & 254.23 GB \\ \hline
         100/140/170/200 & 9,822 & 253.71 GB \\ \hline
         
    \end{tabular}
  \end{center}
\end{table}

This subsection compares the 150/160/190/200 length group to other length groups. Table~\ref{tab:length group comparison} lists the number of usable primers and the obtained tube capacity of different length groups. The scheme used is collisions-based VL-DNA. All length groups are supposed to have a length of 200 so that we can cut the maximum payload length when the DNA sequence has no collisions. We first consider the lengths $\in$ [150,200] to ensure considerable long payload lengths. Meanwhile, the lengths are multiple of 10 for ease of implementation. The first length group in Table~\ref{tab:length group comparison} is our length group. The following eight length groups are all possible groups that have four lengths and the lengths satisfy the above demands. Our length group outperforms all other four length groups in terms of the number of usable primers and obtained tube capacity. That is mainly because the combination of 150/160/190/200 can cover more bases (the number of covered bases is defined in Section.~\ref{sec: variable length}). 150/160/190/200 can have potential cut points every ten bases as early as 450th bases. In comparison, other groups have recurrent cut points since 600+th bases.

We also compare groups that have more than four lengths (the two length groups in the middle). Although the two length groups have five and six variable lengths, they may not be better or just slightly better than the 150/160/190/200 group. Nevertheless, they need one more DNA base to indicate the length used. Considering the overhead of the VL-DNA scheme (i.e., reduction of payload length) is just one or less than one DNA base (Section~\ref{sec:capacity improvement}), trading one more DNA base for the limited increases in the number of covered bases is relatively costly. We only show length groups with up to six lengths because the capacity increment diminishes as the number of lengths increases.

Moreover, we extend the length space from [150,200] to [100,200] to check whether shorter payload lengths can recover significantly more primers (the last two length groups). Since there are two many possible length groups, we only show two of them. Other length groups roughly have similar trends. Shorter payloads do recover more usable primers than our four length group and even the six length group. However, the capability of recovering more primers means cutting more collisions with shorter payloads and thus leading to a suboptimal tube capacity.

The above discussion is based on the current practical payload length (200 bases). However,  the VL-DNA scheme also works well for future DNA strand lengths. As strand length increases, it is reasonable to have more basic lengths to combine more variable lengths because the increased strand length amortizes the overhead of metadata (i.e., indicating length). Moreover, as strand length increases, the capacity of primer also increases. VL-DNA, which targets to recover more usable primers, becomes ever more valuable.

\section{Conclusion}\label{sec:conclusion}
This paper proposes a variable payload length scheme for DNA storage systems. This scheme serves as a post-processing to enhance DNA tube capacity by recovering more usable primers. It can be applied on top of any encoding schemes. Evaluation of multiple existing encoding schemes shows that our scheme can improve tens of Gigabytes of tube capacity, which counts for at least 18.27\% tube capacity.

\begin{acks}
To Robert, for the bagels and explaining CMYK and color spaces.
\end{acks}

\bibliographystyle{ACM-Reference-Format}
\bibliography{ref}


\begin{thebibliography}{104}


\ifx \showCODEN    \undefined \def \showCODEN     #1{\unskip}     \fi
\ifx \showDOI      \undefined \def \showDOI       #1{#1}\fi
\ifx \showISBNx    \undefined \def \showISBNx     #1{\unskip}     \fi
\ifx \showISBNxiii \undefined \def \showISBNxiii  #1{\unskip}     \fi
\ifx \showISSN     \undefined \def \showISSN      #1{\unskip}     \fi
\ifx \showLCCN     \undefined \def \showLCCN      #1{\unskip}     \fi
\ifx \shownote     \undefined \def \shownote      #1{#1}          \fi
\ifx \showarticletitle \undefined \def \showarticletitle #1{#1}   \fi
\ifx \showURL      \undefined \def \showURL       {\relax}        \fi
\providecommand\bibfield[2]{#2}
\providecommand\bibinfo[2]{#2}
\providecommand\natexlab[1]{#1}
\providecommand\showeprint[2][]{arXiv:#2}

\bibitem[com([n.\,d.])]%
        {computationalbiology}
 \bibinfo{year}{[n.\,d.]}\natexlab{}.
\newblock \bibinfo{title}{The digital side of biology.}
\newblock
  \bibinfo{howpublished}{\url{https://phys.org/news/2011-03-digital-side-biology.html}}.
\newblock
\newblock
\shownote{Accessed: 2022-09-12}.


\bibitem[DNA([n.\,d.])]%
        {DNAsimulator}
 \bibinfo{year}{[n.\,d.]}\natexlab{}.
\newblock \bibinfo{title}{DNA Storage Simulation.}
\newblock
  \bibinfo{howpublished}{\url{https://master.dbahb2jho41s4.amplifyapp.com/}}.
\newblock
\newblock
\shownote{Accessed: 2022-09-12}.


\bibitem[Oli([n.\,d.])]%
        {OligoAnalyzer}
 \bibinfo{year}{[n.\,d.]}\natexlab{}.
\newblock \bibinfo{title}{Integrated DNA technology: OligoAnalyzer.}
\newblock \bibinfo{howpublished}{\url{https://www.idtdna.com/calc/analyzer}}.
\newblock
\newblock
\shownote{Accessed: 2022-09-12}.


\bibitem[MSI([n.\,d.])]%
        {MSI}
 \bibinfo{year}{[n.\,d.]}\natexlab{}.
\newblock \bibinfo{title}{Minnesota Supercomputing Institute (MSI) at the
  University of Minnesota.}
\newblock \bibinfo{howpublished}{\url{http://www.msi.umn.edu}}.
\newblock
\newblock
\shownote{Accessed: 2022-09-12}.


\bibitem[del([n.\,d.]a)]%
        {deltaG1}
 \bibinfo{year}{[n.\,d.]}\natexlab{a}.
\newblock \bibinfo{title}{OligoArchitect Online - Glossary of Parameters.}
\newblock
  \bibinfo{howpublished}{\url{https://www.gene-quantification.de/oligo_architect_glossary.pdf}}.
\newblock
\newblock
\shownote{Accessed: 2022-09-12}.


\bibitem[dep([n.\,d.])]%
        {depth}
 \bibinfo{year}{[n.\,d.]}\natexlab{}.
\newblock \bibinfo{title}{Sequencing depth.}
\newblock
  \bibinfo{howpublished}{\url{https://www.genomicseducation.hee.nhs.uk/glossary/read-depth/}}.
\newblock
\newblock
\shownote{Accessed: 2022-09-12}.


\bibitem[del([n.\,d.]b)]%
        {deltaG2}
 \bibinfo{year}{[n.\,d.]}\natexlab{b}.
\newblock \bibinfo{title}{What is delta G value?}
\newblock
  \bibinfo{howpublished}{\url{https://www.researchgate.net/post/What-is-delta-G-value}}.
\newblock
\newblock
\shownote{Accessed: 2022-09-12}.


\bibitem[DNA(2022)]%
        {DNAspeed}
 \bibinfo{year}{\text{[Online] Accessed on Augest. 6, 2022}}\natexlab{}.
\newblock \showarticletitle{New breakthrough gets us closer to using DNA as
  data storage.}
\newblock  (\bibinfo{year}{\text{[Online] Accessed on Augest. 6, 2022}}).
\newblock


\bibitem[Abd-Elsalam(2003)]%
        {abd2003bioinformatic}
\bibfield{author}{\bibinfo{person}{Kamel~A Abd-Elsalam}.}
  \bibinfo{year}{2003}\natexlab{}.
\newblock \showarticletitle{Bioinformatic tools and guideline for PCR primer
  design}.
\newblock \bibinfo{journal}{\emph{african Journal of biotechnology}}
  \bibinfo{volume}{2}, \bibinfo{number}{5} (\bibinfo{year}{2003}),
  \bibinfo{pages}{91--95}.
\newblock


\bibitem[Allentoft et~al\mbox{.}(2012)]%
        {allentoft2012half}
\bibfield{author}{\bibinfo{person}{Morten~E Allentoft},
  \bibinfo{person}{Matthew Collins}, \bibinfo{person}{David Harker},
  \bibinfo{person}{James Haile}, \bibinfo{person}{Charlotte~L Oskam},
  \bibinfo{person}{Marie~L Hale}, \bibinfo{person}{Paula~F Campos},
  \bibinfo{person}{Jose~A Samaniego}, \bibinfo{person}{M~Thomas~P Gilbert},
  \bibinfo{person}{Eske Willerslev}, {et~al\mbox{.}}}
  \bibinfo{year}{2012}\natexlab{}.
\newblock \showarticletitle{The half-life of DNA in bone: measuring decay
  kinetics in 158 dated fossils}.
\newblock \bibinfo{journal}{\emph{Proceedings of the Royal Society B:
  Biological Sciences}} \bibinfo{volume}{279}, \bibinfo{number}{1748}
  (\bibinfo{year}{2012}), \bibinfo{pages}{4724--4733}.
\newblock


\bibitem[Alliance(2021)]%
        {alliance2021preserving}
\bibfield{author}{\bibinfo{person}{DNA Data~Storage Alliance}.}
  \bibinfo{year}{2021}\natexlab{}.
\newblock \bibinfo{booktitle}{\emph{Preserving Our Digital Legacy: an
  Introduction To Dna Data Storage}}.
\newblock \bibinfo{type}{{T}echnical {R}eport}. \bibinfo{institution}{tech.
  rep. June}.
\newblock


\bibitem[Altschul et~al\mbox{.}(1990)]%
        {BLAST}
\bibfield{author}{\bibinfo{person}{Stephen~F Altschul}, \bibinfo{person}{Warren
  Gish}, \bibinfo{person}{Webb Miller}, \bibinfo{person}{Eugene~W Myers}, {and}
  \bibinfo{person}{David~J Lipman}.} \bibinfo{year}{1990}\natexlab{}.
\newblock \showarticletitle{Basic local alignment search tool}.
\newblock \bibinfo{journal}{\emph{Journal of molecular biology}}
  \bibinfo{volume}{215}, \bibinfo{number}{3} (\bibinfo{year}{1990}),
  \bibinfo{pages}{403--410}.
\newblock


\bibitem[Anavy et~al\mbox{.}(2018)]%
        {anavy2018improved}
\bibfield{author}{\bibinfo{person}{Leon Anavy}, \bibinfo{person}{Inbal Vaknin},
  \bibinfo{person}{Orna Atar}, \bibinfo{person}{Roee Amit}, {and}
  \bibinfo{person}{Zohar Yakhini}.} \bibinfo{year}{2018}\natexlab{}.
\newblock \showarticletitle{Improved DNA based storage capacity and fidelity
  using composite DNA letters}.
\newblock \bibinfo{journal}{\emph{bioRxiv}} (\bibinfo{year}{2018}),
  \bibinfo{pages}{433524}.
\newblock


\bibitem[Appuswamy et~al\mbox{.}(2019)]%
        {appuswamy2019oligoarchive}
\bibfield{author}{\bibinfo{person}{Raja Appuswamy}, \bibinfo{person}{Kevin
  Le~Brigand}, \bibinfo{person}{Pascal Barbry}, \bibinfo{person}{Marc
  Antonini}, \bibinfo{person}{Olivier Madderson}, \bibinfo{person}{Paul
  Freemont}, \bibinfo{person}{James McDonald}, {and} \bibinfo{person}{Thomas
  Heinis}.} \bibinfo{year}{2019}\natexlab{}.
\newblock \showarticletitle{OligoArchive: Using DNA in the DBMS storage
  hierarchy.}. In \bibinfo{booktitle}{\emph{CIDR}}.
\newblock


\bibitem[Atikoglu et~al\mbox{.}(2012)]%
        {atikoglu2012workload}
\bibfield{author}{\bibinfo{person}{Berk Atikoglu}, \bibinfo{person}{Yuehai Xu},
  \bibinfo{person}{Eitan Frachtenberg}, \bibinfo{person}{Song Jiang}, {and}
  \bibinfo{person}{Mike Paleczny}.} \bibinfo{year}{2012}\natexlab{}.
\newblock \showarticletitle{Workload analysis of a large-scale key-value
  store}. In \bibinfo{booktitle}{\emph{Proceedings of the 12th ACM
  SIGMETRICS/PERFORMANCE joint international conference on Measurement and
  Modeling of Computer Systems}}. \bibinfo{pages}{53--64}.
\newblock


\bibitem[Batu et~al\mbox{.}(2004)]%
        {batu2004reconstructing}
\bibfield{author}{\bibinfo{person}{Tugkan Batu}, \bibinfo{person}{Sampath
  Kannan}, \bibinfo{person}{Sanjeev Khanna}, {and} \bibinfo{person}{Andrew
  McGregor}.} \bibinfo{year}{2004}\natexlab{}.
\newblock \showarticletitle{Reconstructing strings from random traces}.
\newblock \bibinfo{journal}{\emph{Departmental Papers (CIS)}}
  (\bibinfo{year}{2004}), \bibinfo{pages}{173}.
\newblock


\bibitem[Baum(1995)]%
        {baum1995building}
\bibfield{author}{\bibinfo{person}{Eric~B Baum}.}
  \bibinfo{year}{1995}\natexlab{}.
\newblock \showarticletitle{Building an associative memory vastly larger than
  the brain}.
\newblock \bibinfo{journal}{\emph{Science}} \bibinfo{volume}{268},
  \bibinfo{number}{5210} (\bibinfo{year}{1995}), \bibinfo{pages}{583--585}.
\newblock


\bibitem[Bhagwat et~al\mbox{.}(2009)]%
        {bhagwat2009extreme}
\bibfield{author}{\bibinfo{person}{Deepavali Bhagwat}, \bibinfo{person}{Kave
  Eshghi}, \bibinfo{person}{Darrell~DE Long}, {and} \bibinfo{person}{Mark
  Lillibridge}.} \bibinfo{year}{2009}\natexlab{}.
\newblock \showarticletitle{Extreme binning: Scalable, parallel deduplication
  for chunk-based file backup}. In \bibinfo{booktitle}{\emph{2009 IEEE
  International Symposium on Modeling, Analysis \& Simulation of Computer and
  Telecommunication Systems}}. IEEE, \bibinfo{pages}{1--9}.
\newblock


\bibitem[Blawat et~al\mbox{.}(2016)]%
        {blawat2016forward}
\bibfield{author}{\bibinfo{person}{Meinolf Blawat}, \bibinfo{person}{Klaus
  Gaedke}, \bibinfo{person}{Ingo Huetter}, \bibinfo{person}{Xiao-Ming Chen},
  \bibinfo{person}{Brian Turczyk}, \bibinfo{person}{Samuel Inverso},
  \bibinfo{person}{Benjamin~W Pruitt}, {and} \bibinfo{person}{George~M
  Church}.} \bibinfo{year}{2016}\natexlab{}.
\newblock \showarticletitle{Forward error correction for DNA data storage}.
\newblock \bibinfo{journal}{\emph{Procedia Computer Science}}
  \bibinfo{volume}{80} (\bibinfo{year}{2016}), \bibinfo{pages}{1011--1022}.
\newblock


\bibitem[Bornholt et~al\mbox{.}(2016)]%
        {bornholt2016dna}
\bibfield{author}{\bibinfo{person}{James Bornholt}, \bibinfo{person}{Randolph
  Lopez}, \bibinfo{person}{Douglas~M Carmean}, \bibinfo{person}{Luis Ceze},
  \bibinfo{person}{Georg Seelig}, {and} \bibinfo{person}{Karin Strauss}.}
  \bibinfo{year}{2016}\natexlab{}.
\newblock \showarticletitle{A DNA-based archival storage system}. In
  \bibinfo{booktitle}{\emph{Proceedings of the Twenty-First International
  Conference on Architectural Support for Programming Languages and Operating
  Systems}}. \bibinfo{pages}{637--649}.
\newblock


\bibitem[Cao et~al\mbox{.}(2018)]%
        {cao2018alacc}
\bibfield{author}{\bibinfo{person}{Zhichao Cao}, \bibinfo{person}{Hao Wen},
  \bibinfo{person}{Fenggang Wu}, {and} \bibinfo{person}{David~HC Du}.}
  \bibinfo{year}{2018}\natexlab{}.
\newblock \showarticletitle{$\{$ALACC$\}$: Accelerating restore performance of
  data deduplication systems using adaptive look-ahead window assisted chunk
  caching}. In \bibinfo{booktitle}{\emph{16th $\{$USENIX$\}$ Conference on File
  and Storage Technologies ($\{$FAST$\}$ 18)}}. \bibinfo{pages}{309--324}.
\newblock


\bibitem[Carlson(2017)]%
        {carlson2017guesstimating}
\bibfield{author}{\bibinfo{person}{R Carlson}.}
  \bibinfo{year}{2017}\natexlab{}.
\newblock \showarticletitle{Guesstimating the size of the global array
  synthesis market}.
\newblock \bibinfo{journal}{\emph{Synthesis August}}  \bibinfo{volume}{30}
  (\bibinfo{year}{2017}).
\newblock


\bibitem[Carmean et~al\mbox{.}(2018)]%
        {carmean2018dna}
\bibfield{author}{\bibinfo{person}{Douglas Carmean}, \bibinfo{person}{Luis
  Ceze}, \bibinfo{person}{Georg Seelig}, \bibinfo{person}{Kendall Stewart},
  \bibinfo{person}{Karin Strauss}, {and} \bibinfo{person}{Max Willsey}.}
  \bibinfo{year}{2018}\natexlab{}.
\newblock \showarticletitle{DNA data storage and hybrid molecular--electronic
  computing}.
\newblock \bibinfo{journal}{\emph{Proc. IEEE}} \bibinfo{volume}{107},
  \bibinfo{number}{1} (\bibinfo{year}{2018}), \bibinfo{pages}{63--72}.
\newblock


\bibitem[Ceze et~al\mbox{.}(2019)]%
        {ceze2019molecular}
\bibfield{author}{\bibinfo{person}{Luis Ceze}, \bibinfo{person}{Jeff Nivala},
  {and} \bibinfo{person}{Karin Strauss}.} \bibinfo{year}{2019}\natexlab{}.
\newblock \showarticletitle{Molecular digital data storage using DNA}.
\newblock \bibinfo{journal}{\emph{Nature Reviews Genetics}}
  \bibinfo{volume}{20}, \bibinfo{number}{8} (\bibinfo{year}{2019}),
  \bibinfo{pages}{456--466}.
\newblock


\bibitem[Choi et~al\mbox{.}(2018)]%
        {choi2018addition}
\bibfield{author}{\bibinfo{person}{Yeongjae Choi}, \bibinfo{person}{Taehoon
  Ryu}, \bibinfo{person}{Amos~C Lee}, \bibinfo{person}{Hansol Choi},
  \bibinfo{person}{Hansaem Lee}, \bibinfo{person}{Jaejun Park},
  \bibinfo{person}{Suk-Heung Song}, \bibinfo{person}{Seoju Kim},
  \bibinfo{person}{Hyeli Kim}, \bibinfo{person}{Wook Park}, {et~al\mbox{.}}}
  \bibinfo{year}{2018}\natexlab{}.
\newblock \showarticletitle{Addition of degenerate bases to DNA-based data
  storage for increased information capacity}.
\newblock \bibinfo{journal}{\emph{bioRxiv}} (\bibinfo{year}{2018}),
  \bibinfo{pages}{367052}.
\newblock


\bibitem[Choi et~al\mbox{.}(2019)]%
        {choi2019high}
\bibfield{author}{\bibinfo{person}{Yeongjae Choi}, \bibinfo{person}{Taehoon
  Ryu}, \bibinfo{person}{Amos~C Lee}, \bibinfo{person}{Hansol Choi},
  \bibinfo{person}{Hansaem Lee}, \bibinfo{person}{Jaejun Park},
  \bibinfo{person}{Suk-Heung Song}, \bibinfo{person}{Seojoo Kim},
  \bibinfo{person}{Hyeli Kim}, \bibinfo{person}{Wook Park}, {et~al\mbox{.}}}
  \bibinfo{year}{2019}\natexlab{}.
\newblock \showarticletitle{High information capacity DNA-based data storage
  with augmented encoding characters using degenerate bases}.
\newblock \bibinfo{journal}{\emph{Scientific reports}} \bibinfo{volume}{9},
  \bibinfo{number}{1} (\bibinfo{year}{2019}), \bibinfo{pages}{1--7}.
\newblock


\bibitem[Church et~al\mbox{.}(2012)]%
        {church2012next}
\bibfield{author}{\bibinfo{person}{George~M Church}, \bibinfo{person}{Yuan
  Gao}, {and} \bibinfo{person}{Sriram Kosuri}.}
  \bibinfo{year}{2012}\natexlab{}.
\newblock \showarticletitle{Next-generation digital information storage in
  DNA}.
\newblock \bibinfo{journal}{\emph{Science}} \bibinfo{volume}{337},
  \bibinfo{number}{6102} (\bibinfo{year}{2012}), \bibinfo{pages}{1628--1628}.
\newblock


\bibitem[Corporation(2023)]%
        {IDC}
\bibfield{author}{\bibinfo{person}{International~Data Corporation}.}
  \bibinfo{year}{2023}\natexlab{}.
\newblock \bibinfo{title}{Worldwide Global StorageSphere Forecast, 2021–2025:
  To Save or Not to Save Data, That Is the Question.}
\newblock
\newblock
\urldef\tempurl%
\url{https://www.idc.com/getdoc.jsp?containerId=US47509621}
\showURL{%
Retrieved March 1, 2023 from \tempurl}


\bibitem[Das et~al\mbox{.}(1999)]%
        {das1999studies}
\bibfield{author}{\bibinfo{person}{Simantini Das}, \bibinfo{person}{Satish~C
  Mohapatra}, {and} \bibinfo{person}{James~T Hsu}.}
  \bibinfo{year}{1999}\natexlab{}.
\newblock \showarticletitle{Studies on primer-dimer formation in polymerase
  chain reaction (PCR)}.
\newblock \bibinfo{journal}{\emph{Biotechnology Techniques}}
  \bibinfo{volume}{13}, \bibinfo{number}{10} (\bibinfo{year}{1999}),
  \bibinfo{pages}{643--646}.
\newblock


\bibitem[Deng et~al\mbox{.}(2009)]%
        {imagenet_cvpr09}
\bibfield{author}{\bibinfo{person}{J. Deng}, \bibinfo{person}{W. Dong},
  \bibinfo{person}{R. Socher}, \bibinfo{person}{L.-J. Li}, \bibinfo{person}{K.
  Li}, {and} \bibinfo{person}{L. Fei-Fei}.} \bibinfo{year}{2009}\natexlab{}.
\newblock \showarticletitle{{ImageNet: A Large-Scale Hierarchical Image
  Database}}. In \bibinfo{booktitle}{\emph{CVPR09}}.
\newblock


\bibitem[Deng et~al\mbox{.}(2019)]%
        {deng2019optimized}
\bibfield{author}{\bibinfo{person}{Li Deng}, \bibinfo{person}{Yixin Wang},
  \bibinfo{person}{MD Noor-A-Rahim}, \bibinfo{person}{Yong~Liang Guan},
  \bibinfo{person}{Zhiping Shi}, \bibinfo{person}{Erry Gunawan}, {and}
  \bibinfo{person}{Chueh~Loo Poh}.} \bibinfo{year}{2019}\natexlab{}.
\newblock \showarticletitle{Optimized code design for constrained DNA data
  storage with asymmetric errors}.
\newblock \bibinfo{journal}{\emph{IEEE Access}}  \bibinfo{volume}{7}
  (\bibinfo{year}{2019}), \bibinfo{pages}{84107--84121}.
\newblock


\bibitem[Dieffenbach et~al\mbox{.}(1993)]%
        {dieffenbach1993general}
\bibfield{author}{\bibinfo{person}{CW Dieffenbach}, \bibinfo{person}{TM Lowe},
  {and} \bibinfo{person}{GS Dveksler}.} \bibinfo{year}{1993}\natexlab{}.
\newblock \showarticletitle{General concepts for PCR primer design}.
\newblock \bibinfo{journal}{\emph{PCR methods appl}} \bibinfo{volume}{3},
  \bibinfo{number}{3} (\bibinfo{year}{1993}), \bibinfo{pages}{S30--S37}.
\newblock


\bibitem[Dutch(2008)]%
        {dutch2008understanding}
\bibfield{author}{\bibinfo{person}{Mike Dutch}.}
  \bibinfo{year}{2008}\natexlab{}.
\newblock \showarticletitle{Understanding data deduplication ratios}. In
  \bibinfo{booktitle}{\emph{SNIA Data Management Forum}}. \bibinfo{pages}{7}.
\newblock


\bibitem[El-Shaikh et~al\mbox{.}(2022)]%
        {el2022high}
\bibfield{author}{\bibinfo{person}{Alex El-Shaikh}, \bibinfo{person}{Marius
  Welzel}, \bibinfo{person}{Dominik Heider}, {and} \bibinfo{person}{Bernhard
  Seeger}.} \bibinfo{year}{2022}\natexlab{}.
\newblock \showarticletitle{High-scale random access on DNA storage systems}.
\newblock \bibinfo{journal}{\emph{NAR genomics and bioinformatics}}
  \bibinfo{volume}{4}, \bibinfo{number}{1} (\bibinfo{year}{2022}),
  \bibinfo{pages}{lqab126}.
\newblock


\bibitem[Erlich and Zielinski(2017)]%
        {erlich2017dna}
\bibfield{author}{\bibinfo{person}{Yaniv Erlich} {and} \bibinfo{person}{Dina
  Zielinski}.} \bibinfo{year}{2017}\natexlab{}.
\newblock \showarticletitle{DNA Fountain enables a robust and efficient storage
  architecture}.
\newblock \bibinfo{journal}{\emph{Science}} \bibinfo{volume}{355},
  \bibinfo{number}{6328} (\bibinfo{year}{2017}), \bibinfo{pages}{950--954}.
\newblock


\bibitem[Everitt et~al\mbox{.}([n.\,d.])]%
        {everitt2011cluster}
\bibfield{author}{\bibinfo{person}{Brian~S Everitt}, \bibinfo{person}{Sabine
  Landau}, \bibinfo{person}{Morven Leese}, {and} \bibinfo{person}{Daniel
  Stahl}.} \bibinfo{year}{[n.\,d.]}\natexlab{}.
\newblock \showarticletitle{Cluster Analysis. --John Wiley \& Sons}.
\newblock \bibinfo{journal}{\emph{Ltd., New York}} (\bibinfo{year}{[n.\,d.]}),
  \bibinfo{pages}{330}.
\newblock


\bibitem[Extance(2016)]%
        {extance2016dna}
\bibfield{author}{\bibinfo{person}{Andy Extance}.}
  \bibinfo{year}{2016}\natexlab{}.
\newblock \showarticletitle{How DNA could store all the world's data}.
\newblock \bibinfo{journal}{\emph{Nature}} \bibinfo{volume}{537},
  \bibinfo{number}{7618} (\bibinfo{year}{2016}).
\newblock


\bibitem[Fontana~Jr and Decad(2018)]%
        {fontana2018moore}
\bibfield{author}{\bibinfo{person}{Robert~E Fontana~Jr} {and}
  \bibinfo{person}{Gary~M Decad}.} \bibinfo{year}{2018}\natexlab{}.
\newblock \showarticletitle{Moore’s law realities for recording systems and
  memory storage components: HDD, tape, NAND, and optical}.
\newblock \bibinfo{journal}{\emph{AIP Advances}} \bibinfo{volume}{8},
  \bibinfo{number}{5} (\bibinfo{year}{2018}), \bibinfo{pages}{056506}.
\newblock


\bibitem[for Policy~Studies(2018)]%
        {DNAfuture}
\bibfield{author}{\bibinfo{person}{Potomac~Institute for Policy~Studies}.}
  \bibinfo{year}{2018}\natexlab{}.
\newblock \bibinfo{booktitle}{\emph{The Future of DNA Data Storage}}.
\newblock
\urldef\tempurl%
\url{https://potomacinstitute.org/images/studies/Future_of_DNA_Data_Storage.pdf}
\showURL{%
Retrieved March 1, 2023 from \tempurl}


\bibitem[Fu et~al\mbox{.}(2015)]%
        {fu2015design}
\bibfield{author}{\bibinfo{person}{Min Fu}, \bibinfo{person}{Dan Feng},
  \bibinfo{person}{Yu Hua}, \bibinfo{person}{Xubin He},
  \bibinfo{person}{Zuoning Chen}, \bibinfo{person}{Wen Xia},
  \bibinfo{person}{Yucheng Zhang}, {and} \bibinfo{person}{Yujuan Tan}.}
  \bibinfo{year}{2015}\natexlab{}.
\newblock \showarticletitle{Design tradeoffs for data deduplication performance
  in backup workloads}. In \bibinfo{booktitle}{\emph{13th $\{$USENIX$\}$
  Conference on File and Storage Technologies ($\{$FAST$\}$ 15)}}.
  \bibinfo{pages}{331--344}.
\newblock


\bibitem[Gantz and Reinsel(2010)]%
        {IDCdata2010}
\bibfield{author}{\bibinfo{person}{John Gantz} {and} \bibinfo{person}{David
  Reinsel}.} \bibinfo{year}{2010}\natexlab{}.
\newblock \showarticletitle{The Digital Universe Decade – Are You Ready?}
\newblock \bibinfo{journal}{\emph{IDC White Paper}} (\bibinfo{year}{2010}).
\newblock


\bibitem[Goldman et~al\mbox{.}(2013)]%
        {goldman2013towards}
\bibfield{author}{\bibinfo{person}{Nick Goldman}, \bibinfo{person}{Paul
  Bertone}, \bibinfo{person}{Siyuan Chen}, \bibinfo{person}{Christophe
  Dessimoz}, \bibinfo{person}{Emily~M LeProust}, \bibinfo{person}{Botond
  Sipos}, {and} \bibinfo{person}{Ewan Birney}.}
  \bibinfo{year}{2013}\natexlab{}.
\newblock \showarticletitle{Towards practical, high-capacity, low-maintenance
  information storage in synthesized DNA}.
\newblock \bibinfo{journal}{\emph{Nature}} \bibinfo{volume}{494},
  \bibinfo{number}{7435} (\bibinfo{year}{2013}), \bibinfo{pages}{77--80}.
\newblock


\bibitem[Grass et~al\mbox{.}(2015)]%
        {grass2015robust}
\bibfield{author}{\bibinfo{person}{Robert~N Grass}, \bibinfo{person}{Reinhard
  Heckel}, \bibinfo{person}{Michela Puddu}, \bibinfo{person}{Daniela Paunescu},
  {and} \bibinfo{person}{Wendelin~J Stark}.} \bibinfo{year}{2015}\natexlab{}.
\newblock \showarticletitle{Robust chemical preservation of digital information
  on DNA in silica with error-correcting codes}.
\newblock \bibinfo{journal}{\emph{Angewandte Chemie International Edition}}
  \bibinfo{volume}{54}, \bibinfo{number}{8} (\bibinfo{year}{2015}),
  \bibinfo{pages}{2552--2555}.
\newblock


\bibitem[Guo and Efstathopoulos(2011)]%
        {guo2011building}
\bibfield{author}{\bibinfo{person}{Fanglu Guo} {and} \bibinfo{person}{Petros
  Efstathopoulos}.} \bibinfo{year}{2011}\natexlab{}.
\newblock \showarticletitle{Building a High-performance Deduplication System.}.
  In \bibinfo{booktitle}{\emph{USENIX annual technical conference}}.
\newblock


\bibitem[Heinis(2019)]%
        {heinis2019survey}
\bibfield{author}{\bibinfo{person}{Thomas Heinis}.}
  \bibinfo{year}{2019}\natexlab{}.
\newblock \showarticletitle{Survey of Information Encoding Techniques for DNA}.
\newblock \bibinfo{journal}{\emph{arXiv preprint arXiv:1906.11062}}
  (\bibinfo{year}{2019}).
\newblock


\bibitem[InternetArchive(2023)]%
        {InternetArchive}
\bibfield{author}{\bibinfo{person}{InternetArchive}.}
  \bibinfo{year}{2023}\natexlab{}.
\newblock \bibinfo{title}{Internet Archive Public library.}
\newblock
\newblock
\urldef\tempurl%
\url{https://archive.org/}
\showURL{%
Retrieved March 1, 2023 from \tempurl}


\bibitem[Jain et~al\mbox{.}(2020)]%
        {Jain2020CodingFO}
\bibfield{author}{\bibinfo{person}{Siddhartha Jain}, \bibinfo{person}{Farzad
  Farnoud}, \bibinfo{person}{Moshe Schwartz}, {and} \bibinfo{person}{Jehoshua
  Bruck}.} \bibinfo{year}{2020}\natexlab{}.
\newblock \showarticletitle{Coding for Optimized Writing Rate in DNA Storage}.
\newblock \bibinfo{journal}{\emph{2020 IEEE International Symposium on
  Information Theory (ISIT)}} (\bibinfo{year}{2020}),
  \bibinfo{pages}{711--716}.
\newblock


\bibitem[Kannan and Zacharias(2007)]%
        {kannan2007folding}
\bibfield{author}{\bibinfo{person}{Srinivasaraghavan Kannan} {and}
  \bibinfo{person}{Martin Zacharias}.} \bibinfo{year}{2007}\natexlab{}.
\newblock \showarticletitle{Folding of a DNA hairpin loop structure in explicit
  solvent using replica-exchange molecular dynamics simulations}.
\newblock \bibinfo{journal}{\emph{Biophysical journal}} \bibinfo{volume}{93},
  \bibinfo{number}{9} (\bibinfo{year}{2007}), \bibinfo{pages}{3218--3228}.
\newblock


\bibitem[Korte and Vygen(2012)]%
        {korte2012bin}
\bibfield{author}{\bibinfo{person}{Bernhard Korte} {and} \bibinfo{person}{Jens
  Vygen}.} \bibinfo{year}{2012}\natexlab{}.
\newblock \showarticletitle{Bin-packing}.
\newblock In \bibinfo{booktitle}{\emph{Kombinatorische Optimierung}}.
  \bibinfo{publisher}{Springer}, \bibinfo{pages}{499--516}.
\newblock


\bibitem[Kosuri and Church(2014)]%
        {kosuri2014large}
\bibfield{author}{\bibinfo{person}{Sriram Kosuri} {and}
  \bibinfo{person}{George~M Church}.} \bibinfo{year}{2014}\natexlab{}.
\newblock \showarticletitle{Large-scale de novo DNA synthesis: technologies and
  applications}.
\newblock \bibinfo{journal}{\emph{Nature methods}} \bibinfo{volume}{11},
  \bibinfo{number}{5} (\bibinfo{year}{2014}), \bibinfo{pages}{499}.
\newblock


\bibitem[Kruus et~al\mbox{.}(2010)]%
        {kruus2010bimodal}
\bibfield{author}{\bibinfo{person}{Erik Kruus}, \bibinfo{person}{Cristian
  Ungureanu}, {and} \bibinfo{person}{Cezary Dubnicki}.}
  \bibinfo{year}{2010}\natexlab{}.
\newblock \showarticletitle{Bimodal content defined chunking for backup
  streams.}. In \bibinfo{booktitle}{\emph{Fast}}. \bibinfo{pages}{239--252}.
\newblock


\bibitem[Lai et~al\mbox{.}(2015)]%
        {lai2015atlas}
\bibfield{author}{\bibinfo{person}{Chunbo Lai}, \bibinfo{person}{Song Jiang},
  \bibinfo{person}{Liqiong Yang}, \bibinfo{person}{Shiding Lin},
  \bibinfo{person}{Guangyu Sun}, \bibinfo{person}{Zhenyu Hou},
  \bibinfo{person}{Can Cui}, {and} \bibinfo{person}{Jason Cong}.}
  \bibinfo{year}{2015}\natexlab{}.
\newblock \showarticletitle{Atlas: Baidu's key-value storage system for cloud
  data}. In \bibinfo{booktitle}{\emph{2015 31st Symposium on Mass Storage
  Systems and Technologies (MSST)}}. IEEE, \bibinfo{pages}{1--14}.
\newblock


\bibitem[Lee et~al\mbox{.}(2018)]%
        {lee2018enzymatic}
\bibfield{author}{\bibinfo{person}{Henry~H Lee}, \bibinfo{person}{Reza Kalhor},
  \bibinfo{person}{Naveen Goela}, \bibinfo{person}{Jean Bolot}, {and}
  \bibinfo{person}{George~M Church}.} \bibinfo{year}{2018}\natexlab{}.
\newblock \showarticletitle{Enzymatic DNA synthesis for digital information
  storage}.
\newblock \bibinfo{journal}{\emph{bioRxiv}} (\bibinfo{year}{2018}),
  \bibinfo{pages}{348987}.
\newblock


\bibitem[Lenz et~al\mbox{.}(2020)]%
        {Lenz2020CodingOS}
\bibfield{author}{\bibinfo{person}{Andreas Lenz}, \bibinfo{person}{Paul~H.
  Siegel}, \bibinfo{person}{Antonia Wachter-Zeh}, {and} \bibinfo{person}{Eitan
  Yaakobi}.} \bibinfo{year}{2020}\natexlab{}.
\newblock \showarticletitle{Coding Over Sets for DNA Storage}.
\newblock \bibinfo{journal}{\emph{IEEE Transactions on Information Theory}}
  \bibinfo{volume}{66} (\bibinfo{year}{2020}), \bibinfo{pages}{2331--2351}.
\newblock


\bibitem[Li et~al\mbox{.}(2021)]%
        {li2021img}
\bibfield{author}{\bibinfo{person}{Bingzhe Li}, \bibinfo{person}{Li Ou}, {and}
  \bibinfo{person}{David Du}.} \bibinfo{year}{2021}\natexlab{}.
\newblock \showarticletitle{Img-dna: approximate dna storage for images}. In
  \bibinfo{booktitle}{\emph{Proceedings of the 14th ACM International
  Conference on Systems and Storage}}. \bibinfo{pages}{1--9}.
\newblock


\bibitem[Li et~al\mbox{.}(2020)]%
        {li2020can}
\bibfield{author}{\bibinfo{person}{Bingzhe Li}, \bibinfo{person}{Nae~Young
  Song}, \bibinfo{person}{Li Ou}, {and} \bibinfo{person}{David~HC Du}.}
  \bibinfo{year}{2020}\natexlab{}.
\newblock \showarticletitle{Can We Store the Whole World's Data in $\{$DNA$\}$
  Storage?}. In \bibinfo{booktitle}{\emph{12th $\{$USENIX$\}$ Workshop on Hot
  Topics in Storage and File Systems (HotStorage 20)}}.
\newblock


\bibitem[Lin et~al\mbox{.}(2022)]%
        {lin2022managing}
\bibfield{author}{\bibinfo{person}{Dehui Lin}, \bibinfo{person}{Yasamin
  Tabatabaee}, \bibinfo{person}{Yash Pote}, {and} \bibinfo{person}{Djordje
  Jevdjic}.} \bibinfo{year}{2022}\natexlab{}.
\newblock \showarticletitle{Managing reliability skew in DNA storage}. In
  \bibinfo{booktitle}{\emph{Proceedings of the 49th Annual International
  Symposium on Computer Architecture}}. \bibinfo{pages}{482--494}.
\newblock


\bibitem[Lu and Kim(2021)]%
        {Lu2021DesignON}
\bibfield{author}{\bibinfo{person}{Xiaozhou Lu} {and} \bibinfo{person}{Sunghwan
  Kim}.} \bibinfo{year}{2021}\natexlab{}.
\newblock \showarticletitle{Design of Nonbinary Error Correction Codes With a
  Maximum Run-Length Constraint to Correct a Single Insertion or Deletion Error
  for DNA Storage}.
\newblock \bibinfo{journal}{\emph{IEEE Access}}  \bibinfo{volume}{9}
  (\bibinfo{year}{2021}), \bibinfo{pages}{135354--135363}.
\newblock


\bibitem[Luby(2002)]%
        {luby2002lt}
\bibfield{author}{\bibinfo{person}{Michael Luby}.}
  \bibinfo{year}{2002}\natexlab{}.
\newblock \showarticletitle{LT codes}. In \bibinfo{booktitle}{\emph{The 43rd
  Annual IEEE Symposium on Foundations of Computer Science, 2002.
  Proceedings.}} IEEE Computer Society, \bibinfo{pages}{271--271}.
\newblock


\bibitem[Ma et~al\mbox{.}(2006)]%
        {ma2006exploring}
\bibfield{author}{\bibinfo{person}{Hairong Ma}, \bibinfo{person}{David~J
  Proctor}, \bibinfo{person}{Elzbieta Kierzek}, \bibinfo{person}{Ryszard
  Kierzek}, \bibinfo{person}{Philip~C Bevilacqua}, {and}
  \bibinfo{person}{Martin Gruebele}.} \bibinfo{year}{2006}\natexlab{}.
\newblock \showarticletitle{Exploring the energy landscape of a small RNA
  hairpin}.
\newblock \bibinfo{journal}{\emph{Journal of the American Chemical Society}}
  \bibinfo{volume}{128}, \bibinfo{number}{5} (\bibinfo{year}{2006}),
  \bibinfo{pages}{1523--1530}.
\newblock


\bibitem[Matange et~al\mbox{.}(2021)]%
        {matange2021dna}
\bibfield{author}{\bibinfo{person}{Karishma Matange}, \bibinfo{person}{James~M
  Tuck}, {and} \bibinfo{person}{Albert~J Keung}.}
  \bibinfo{year}{2021}\natexlab{}.
\newblock \showarticletitle{DNA stability: a central design consideration for
  DNA data storage systems}.
\newblock \bibinfo{journal}{\emph{Nature communications}} \bibinfo{volume}{12},
  \bibinfo{number}{1} (\bibinfo{year}{2021}), \bibinfo{pages}{1--9}.
\newblock


\bibitem[Matteucci and Caruthers(1981)]%
        {matteucci1981synthesis}
\bibfield{author}{\bibinfo{person}{Mark~Douglas Matteucci} {and}
  \bibinfo{person}{M~Ho Caruthers}.} \bibinfo{year}{1981}\natexlab{}.
\newblock \showarticletitle{Synthesis of deoxyoligonucleotides on a polymer
  support}.
\newblock \bibinfo{journal}{\emph{Journal of the American Chemical Society}}
  \bibinfo{volume}{103}, \bibinfo{number}{11} (\bibinfo{year}{1981}),
  \bibinfo{pages}{3185--3191}.
\newblock


\bibitem[Meyer and Bolosky(2012)]%
        {meyer2012study}
\bibfield{author}{\bibinfo{person}{Dutch~T Meyer} {and}
  \bibinfo{person}{William~J Bolosky}.} \bibinfo{year}{2012}\natexlab{}.
\newblock \showarticletitle{A study of practical deduplication}.
\newblock \bibinfo{journal}{\emph{ACM Transactions on Storage (ToS)}}
  \bibinfo{volume}{7}, \bibinfo{number}{4} (\bibinfo{year}{2012}),
  \bibinfo{pages}{1--20}.
\newblock


\bibitem[Miller(2020)]%
        {miller2020future}
\bibfield{author}{\bibinfo{person}{Ethan~L Miller}.}
  \bibinfo{year}{2020}\natexlab{}.
\newblock \showarticletitle{The Future of the Past: Challenges in Archival
  Storage}.
\newblock  (\bibinfo{year}{2020}).
\newblock


\bibitem[Muthitacharoen et~al\mbox{.}(2001)]%
        {muthitacharoen2001low}
\bibfield{author}{\bibinfo{person}{Athicha Muthitacharoen},
  \bibinfo{person}{Benjie Chen}, {and} \bibinfo{person}{David Mazieres}.}
  \bibinfo{year}{2001}\natexlab{}.
\newblock \showarticletitle{A low-bandwidth network file system}. In
  \bibinfo{booktitle}{\emph{Proceedings of the eighteenth ACM symposium on
  Operating systems principles}}. \bibinfo{pages}{174--187}.
\newblock


\bibitem[Nemhauser et~al\mbox{.}(1978)]%
        {nemhauser1978analysis}
\bibfield{author}{\bibinfo{person}{George~L Nemhauser},
  \bibinfo{person}{Laurence~A Wolsey}, {and} \bibinfo{person}{Marshall~L
  Fisher}.} \bibinfo{year}{1978}\natexlab{}.
\newblock \showarticletitle{An analysis of approximations for maximizing
  submodular set functions—I}.
\newblock \bibinfo{journal}{\emph{Mathematical programming}}
  \bibinfo{volume}{14}, \bibinfo{number}{1} (\bibinfo{year}{1978}),
  \bibinfo{pages}{265--294}.
\newblock


\bibitem[Newman(2004)]%
        {newman2004fast}
\bibfield{author}{\bibinfo{person}{Mark~EJ Newman}.}
  \bibinfo{year}{2004}\natexlab{}.
\newblock \showarticletitle{Fast algorithm for detecting community structure in
  networks}.
\newblock \bibinfo{journal}{\emph{Physical review E}} \bibinfo{volume}{69},
  \bibinfo{number}{6} (\bibinfo{year}{2004}), \bibinfo{pages}{066133}.
\newblock


\bibitem[Newman and Girvan(2004)]%
        {newman2004finding}
\bibfield{author}{\bibinfo{person}{Mark~EJ Newman} {and}
  \bibinfo{person}{Michelle Girvan}.} \bibinfo{year}{2004}\natexlab{}.
\newblock \showarticletitle{Finding and evaluating community structure in
  networks}.
\newblock \bibinfo{journal}{\emph{Physical review E}} \bibinfo{volume}{69},
  \bibinfo{number}{2} (\bibinfo{year}{2004}), \bibinfo{pages}{026113}.
\newblock


\bibitem[Newman et~al\mbox{.}(2019)]%
        {newman2019high}
\bibfield{author}{\bibinfo{person}{Sharon Newman}, \bibinfo{person}{Ashley~P
  Stephenson}, \bibinfo{person}{Max Willsey}, \bibinfo{person}{Bichlien~H
  Nguyen}, \bibinfo{person}{Christopher~N Takahashi}, \bibinfo{person}{Karin
  Strauss}, {and} \bibinfo{person}{Luis Ceze}.}
  \bibinfo{year}{2019}\natexlab{}.
\newblock \showarticletitle{High density DNA data storage library via
  dehydration with digital microfluidic retrieval}.
\newblock \bibinfo{journal}{\emph{Nature communications}} \bibinfo{volume}{10},
  \bibinfo{number}{1} (\bibinfo{year}{2019}), \bibinfo{pages}{1--6}.
\newblock


\bibitem[Niedringhaus et~al\mbox{.}(2011)]%
        {niedringhaus2011landscape}
\bibfield{author}{\bibinfo{person}{Thomas~P Niedringhaus},
  \bibinfo{person}{Denitsa Milanova}, \bibinfo{person}{Matthew~B Kerby},
  \bibinfo{person}{Michael~P Snyder}, {and} \bibinfo{person}{Annelise~E
  Barron}.} \bibinfo{year}{2011}\natexlab{}.
\newblock \showarticletitle{Landscape of next-generation sequencing
  technologies}.
\newblock \bibinfo{journal}{\emph{Analytical chemistry}} \bibinfo{volume}{83},
  \bibinfo{number}{12} (\bibinfo{year}{2011}), \bibinfo{pages}{4327--4341}.
\newblock


\bibitem[Organick et~al\mbox{.}(2018)]%
        {organick2018random}
\bibfield{author}{\bibinfo{person}{Lee Organick}, \bibinfo{person}{Siena~Dumas
  Ang}, \bibinfo{person}{Yuan-Jyue Chen}, \bibinfo{person}{Randolph Lopez},
  \bibinfo{person}{Sergey Yekhanin}, \bibinfo{person}{Konstantin Makarychev},
  \bibinfo{person}{Miklos~Z Racz}, \bibinfo{person}{Govinda Kamath},
  \bibinfo{person}{Parikshit Gopalan}, \bibinfo{person}{Bichlien Nguyen},
  {et~al\mbox{.}}} \bibinfo{year}{2018}\natexlab{}.
\newblock \showarticletitle{Random access in large-scale DNA data storage}.
\newblock \bibinfo{journal}{\emph{Nature biotechnology}} \bibinfo{volume}{36},
  \bibinfo{number}{3} (\bibinfo{year}{2018}), \bibinfo{pages}{242}.
\newblock


\bibitem[Organick et~al\mbox{.}(2020)]%
        {organick2020probing}
\bibfield{author}{\bibinfo{person}{Lee Organick}, \bibinfo{person}{Yuan-Jyue
  Chen}, \bibinfo{person}{Siena~Dumas Ang}, \bibinfo{person}{Randolph Lopez},
  \bibinfo{person}{Xiaomeng Liu}, \bibinfo{person}{Karin Strauss}, {and}
  \bibinfo{person}{Luis Ceze}.} \bibinfo{year}{2020}\natexlab{}.
\newblock \showarticletitle{Probing the physical limits of reliable DNA data
  retrieval}.
\newblock \bibinfo{journal}{\emph{Nature communications}} \bibinfo{volume}{11},
  \bibinfo{number}{1} (\bibinfo{year}{2020}), \bibinfo{pages}{1--7}.
\newblock


\bibitem[Panayotov et~al\mbox{.}(2015)]%
        {panayotov2015librispeech}
\bibfield{author}{\bibinfo{person}{Vassil Panayotov}, \bibinfo{person}{Guoguo
  Chen}, \bibinfo{person}{Daniel Povey}, {and} \bibinfo{person}{Sanjeev
  Khudanpur}.} \bibinfo{year}{2015}\natexlab{}.
\newblock \showarticletitle{Librispeech: an asr corpus based on public domain
  audio books}. In \bibinfo{booktitle}{\emph{2015 IEEE international conference
  on acoustics, speech and signal processing (ICASSP)}}. IEEE,
  \bibinfo{pages}{5206--5210}.
\newblock


\bibitem[Popescu et~al\mbox{.}(2012)]%
        {popescu2012same}
\bibfield{author}{\bibinfo{person}{Adrian~Daniel Popescu}, \bibinfo{person}{Vuk
  Ercegovac}, \bibinfo{person}{Andrey Balmin}, \bibinfo{person}{Miguel Branco},
  {and} \bibinfo{person}{Anastasia Ailamaki}.} \bibinfo{year}{2012}\natexlab{}.
\newblock \showarticletitle{Same queries, different data: Can we predict
  runtime performance?}. In \bibinfo{booktitle}{\emph{2012 IEEE 28th
  International Conference on Data Engineering Workshops}}. IEEE,
  \bibinfo{pages}{275--280}.
\newblock


\bibitem[Quinlan and Dorward(2002)]%
        {quinlan2002venti}
\bibfield{author}{\bibinfo{person}{Sean Quinlan} {and} \bibinfo{person}{Sean
  Dorward}.} \bibinfo{year}{2002}\natexlab{}.
\newblock \showarticletitle{Venti: A New Approach to Archival Storage.}. In
  \bibinfo{booktitle}{\emph{FAST}}, Vol.~\bibinfo{volume}{2}.
  \bibinfo{pages}{89--101}.
\newblock


\bibitem[Reinsel et~al\mbox{.}(2018)]%
        {reinsel2018digitization}
\bibfield{author}{\bibinfo{person}{David Reinsel}, \bibinfo{person}{John
  Gantz}, {and} \bibinfo{person}{John Rydning}.}
  \bibinfo{year}{2018}\natexlab{}.
\newblock \showarticletitle{The digitization of the world from edge to core}.
\newblock \bibinfo{journal}{\emph{IDC White Paper}} (\bibinfo{year}{2018}).
\newblock


\bibitem[Richterich(1998)]%
        {richterich1998estimation}
\bibfield{author}{\bibinfo{person}{Peter Richterich}.}
  \bibinfo{year}{1998}\natexlab{}.
\newblock \showarticletitle{Estimation of errors in “raw” DNA sequences: a
  validation study}.
\newblock \bibinfo{journal}{\emph{Genome Research}} \bibinfo{volume}{8},
  \bibinfo{number}{3} (\bibinfo{year}{1998}), \bibinfo{pages}{251--259}.
\newblock


\bibitem[Rutten et~al\mbox{.}(2018)]%
        {rutten2018encoding}
\bibfield{author}{\bibinfo{person}{Martin~GTA Rutten}, \bibinfo{person}{Frits~W
  Vaandrager}, \bibinfo{person}{Johannes~AAW Elemans}, {and}
  \bibinfo{person}{Roeland~JM Nolte}.} \bibinfo{year}{2018}\natexlab{}.
\newblock \showarticletitle{Encoding information into polymers}.
\newblock \bibinfo{journal}{\emph{Nature Reviews Chemistry}}
  \bibinfo{volume}{2}, \bibinfo{number}{11} (\bibinfo{year}{2018}),
  \bibinfo{pages}{365--381}.
\newblock


\bibitem[Sella et~al\mbox{.}(2021)]%
        {sella2021dna}
\bibfield{author}{\bibinfo{person}{Omer~S Sella}, \bibinfo{person}{Amir
  Apelbaum}, \bibinfo{person}{Thomas Heinis}, \bibinfo{person}{Jasmine Quah},
  {and} \bibinfo{person}{Andrew~W Moore}.} \bibinfo{year}{2021}\natexlab{}.
\newblock \showarticletitle{DNA archival storage, a bottom up approach}. In
  \bibinfo{booktitle}{\emph{Proceedings of the 13th ACM Workshop on Hot Topics
  in Storage and File Systems}}. \bibinfo{pages}{58--63}.
\newblock


\bibitem[Singh et~al\mbox{.}(2000)]%
        {singh2000effect}
\bibfield{author}{\bibinfo{person}{Vinay~K Singh}, \bibinfo{person}{R
  Govindarajan}, \bibinfo{person}{Sita Naik}, {and} \bibinfo{person}{Anil
  Kumar}.} \bibinfo{year}{2000}\natexlab{}.
\newblock \showarticletitle{The effect of hairpin structure on PCR
  amplification efficiency}.
\newblock \bibinfo{journal}{\emph{Mol Biol Today}} \bibinfo{volume}{1},
  \bibinfo{number}{3} (\bibinfo{year}{2000}), \bibinfo{pages}{67--69}.
\newblock


\bibitem[Song et~al\mbox{.}(2019)]%
        {song2019multidimensional}
\bibfield{author}{\bibinfo{person}{Xin Song}, \bibinfo{person}{Shalin Shah},
  {and} \bibinfo{person}{John Reif}.} \bibinfo{year}{2019}\natexlab{}.
\newblock \showarticletitle{Multidimensional data organization and random
  access in large-scale DNA storage systems}.
\newblock \bibinfo{journal}{\emph{bioRxiv}} (\bibinfo{year}{2019}),
  \bibinfo{pages}{743369}.
\newblock


\bibitem[Steadman and Fair(2012)]%
        {steadman2012variable}
\bibfield{author}{\bibinfo{person}{Andrew Steadman} {and} \bibinfo{person}{Ivan
  Fair}.} \bibinfo{year}{2012}\natexlab{}.
\newblock \showarticletitle{Variable-length constrained sequence codes}.
\newblock \bibinfo{journal}{\emph{IEEE communications letters}}
  \bibinfo{volume}{17}, \bibinfo{number}{1} (\bibinfo{year}{2012}),
  \bibinfo{pages}{139--142}.
\newblock


\bibitem[Steadman and Fair(2016)]%
        {steadman2016simplified}
\bibfield{author}{\bibinfo{person}{Andrew Steadman} {and} \bibinfo{person}{Ivan
  Fair}.} \bibinfo{year}{2016}\natexlab{}.
\newblock \showarticletitle{Simplified search and construction of
  capacity-approaching variable-length constrained sequence codes}.
\newblock \bibinfo{journal}{\emph{IET Communications}} \bibinfo{volume}{10},
  \bibinfo{number}{14} (\bibinfo{year}{2016}), \bibinfo{pages}{1697--1704}.
\newblock


\bibitem[Tarasov et~al\mbox{.}(2012)]%
        {tarasov2012generating}
\bibfield{author}{\bibinfo{person}{Vasily Tarasov}, \bibinfo{person}{Amar
  Mudrankit}, \bibinfo{person}{Will Buik}, \bibinfo{person}{Philip Shilane},
  \bibinfo{person}{Geoff Kuenning}, {and} \bibinfo{person}{Erez Zadok}.}
  \bibinfo{year}{2012}\natexlab{}.
\newblock \showarticletitle{Generating realistic datasets for deduplication
  analysis}. In \bibinfo{booktitle}{\emph{Presented as part of the 2012 USENIX
  Annual Technical Conference (ATC 12)}}. \bibinfo{pages}{261--272}.
\newblock


\bibitem[Tomek et~al\mbox{.}(2019)]%
        {tomek2019driving}
\bibfield{author}{\bibinfo{person}{Kyle~J Tomek}, \bibinfo{person}{Kevin
  Volkel}, \bibinfo{person}{Alexander Simpson}, \bibinfo{person}{Austin~G
  Hass}, \bibinfo{person}{Elaine~W Indermaur}, \bibinfo{person}{James~M Tuck},
  {and} \bibinfo{person}{Albert~J Keung}.} \bibinfo{year}{2019}\natexlab{}.
\newblock \showarticletitle{Driving the scalability of DNA-based information
  storage systems}.
\newblock \bibinfo{journal}{\emph{ACS synthetic biology}} \bibinfo{volume}{8},
  \bibinfo{number}{6} (\bibinfo{year}{2019}), \bibinfo{pages}{1241--1248}.
\newblock


\bibitem[Van~Dijk et~al\mbox{.}(2014)]%
        {van2014ten}
\bibfield{author}{\bibinfo{person}{Erwin~L Van~Dijk},
  \bibinfo{person}{H{\'e}l{\`e}ne Auger}, \bibinfo{person}{Yan Jaszczyszyn},
  {and} \bibinfo{person}{Claude Thermes}.} \bibinfo{year}{2014}\natexlab{}.
\newblock \showarticletitle{Ten years of next-generation sequencing
  technology}.
\newblock \bibinfo{journal}{\emph{Trends in genetics}} \bibinfo{volume}{30},
  \bibinfo{number}{9} (\bibinfo{year}{2014}), \bibinfo{pages}{418--426}.
\newblock


\bibitem[Wallace et~al\mbox{.}(2012)]%
        {wallace2012characteristics}
\bibfield{author}{\bibinfo{person}{Grant Wallace}, \bibinfo{person}{Fred
  Douglis}, \bibinfo{person}{Hangwei Qian}, \bibinfo{person}{Philip Shilane},
  \bibinfo{person}{Stephen Smaldone}, \bibinfo{person}{Mark Chamness}, {and}
  \bibinfo{person}{Windsor Hsu}.} \bibinfo{year}{2012}\natexlab{}.
\newblock \showarticletitle{Characteristics of backup workloads in production
  systems.}. In \bibinfo{booktitle}{\emph{FAST}}, Vol.~\bibinfo{volume}{12}.
  \bibinfo{pages}{4--4}.
\newblock


\bibitem[Wang et~al\mbox{.}(2019)]%
        {wang2019high}
\bibfield{author}{\bibinfo{person}{Yixin Wang}, \bibinfo{person}{Md
  Noor-A-Rahim}, \bibinfo{person}{Jingyun Zhang}, \bibinfo{person}{Erry
  Gunawan}, \bibinfo{person}{Yong~Liang Guan}, {and} \bibinfo{person}{Chueh~Loo
  Poh}.} \bibinfo{year}{2019}\natexlab{}.
\newblock \showarticletitle{High capacity DNA data storage with variable-length
  Oligonucleotides using repeat accumulate code and hybrid mapping}.
\newblock \bibinfo{journal}{\emph{Journal of biological engineering}}
  \bibinfo{volume}{13}, \bibinfo{number}{1} (\bibinfo{year}{2019}),
  \bibinfo{pages}{1--11}.
\newblock


\bibitem[Wei et~al\mbox{.}(2022)]%
        {wei2022dna}
\bibfield{author}{\bibinfo{person}{Yixun Wei}, \bibinfo{person}{Bingzhe Li},
  {and} \bibinfo{person}{David~HC Du}.} \bibinfo{year}{2022}\natexlab{}.
\newblock \showarticletitle{DNA Storage: A Promising Large Scale Archival
  Storage?}
\newblock \bibinfo{journal}{\emph{arXiv preprint arXiv:2204.01870}}
  (\bibinfo{year}{2022}).
\newblock


\bibitem[Wu et~al\mbox{.}(2017)]%
        {wu2017cost}
\bibfield{author}{\bibinfo{person}{Jie Wu}, \bibinfo{person}{Yu Hua},
  \bibinfo{person}{Pengfei Zuo}, {and} \bibinfo{person}{Yuanyuan Sun}.}
  \bibinfo{year}{2017}\natexlab{}.
\newblock \showarticletitle{A cost-efficient rewriting scheme to improve
  restore performance in deduplication systems}. In
  \bibinfo{booktitle}{\emph{Proc. MSST}}.
\newblock


\bibitem[Wu et~al\mbox{.}(2021)]%
        {Wu2021HDCodeEH}
\bibfield{author}{\bibinfo{person}{Jianjun Wu}, \bibinfo{person}{Shufang
  Zhang}, \bibinfo{person}{Tao Zhang}, {and} \bibinfo{person}{Yuhong Liu}.}
  \bibinfo{year}{2021}\natexlab{}.
\newblock \showarticletitle{HD-Code: End-to-End High Density Code for DNA
  Storage}.
\newblock \bibinfo{journal}{\emph{IEEE Transactions on NanoBioscience}}
  \bibinfo{volume}{20} (\bibinfo{year}{2021}), \bibinfo{pages}{455--463}.
\newblock


\bibitem[Xia et~al\mbox{.}(2016)]%
        {xia2016comprehensive}
\bibfield{author}{\bibinfo{person}{Wen Xia}, \bibinfo{person}{Hong Jiang},
  \bibinfo{person}{Dan Feng}, \bibinfo{person}{Fred Douglis},
  \bibinfo{person}{Philip Shilane}, \bibinfo{person}{Yu Hua},
  \bibinfo{person}{Min Fu}, \bibinfo{person}{Yucheng Zhang}, {and}
  \bibinfo{person}{Yukun Zhou}.} \bibinfo{year}{2016}\natexlab{}.
\newblock \showarticletitle{A comprehensive study of the past, present, and
  future of data deduplication}.
\newblock \bibinfo{journal}{\emph{Proc. IEEE}} \bibinfo{volume}{104},
  \bibinfo{number}{9} (\bibinfo{year}{2016}), \bibinfo{pages}{1681--1710}.
\newblock


\bibitem[Xia et~al\mbox{.}(2014a)]%
        {xia2014similarity}
\bibfield{author}{\bibinfo{person}{Wen Xia}, \bibinfo{person}{Hong Jiang},
  \bibinfo{person}{Dan Feng}, {and} \bibinfo{person}{Yu Hua}.}
  \bibinfo{year}{2014}\natexlab{a}.
\newblock \showarticletitle{Similarity and locality based indexing for high
  performance data deduplication}.
\newblock \bibinfo{journal}{\emph{IEEE transactions on computers}}
  \bibinfo{volume}{64}, \bibinfo{number}{4} (\bibinfo{year}{2014}),
  \bibinfo{pages}{1162--1176}.
\newblock


\bibitem[Xia et~al\mbox{.}(2014b)]%
        {xia2014combining}
\bibfield{author}{\bibinfo{person}{Wen Xia}, \bibinfo{person}{Hong Jiang},
  \bibinfo{person}{Dan Feng}, {and} \bibinfo{person}{Lei Tian}.}
  \bibinfo{year}{2014}\natexlab{b}.
\newblock \showarticletitle{Combining deduplication and delta compression to
  achieve low-overhead data reduction on backup datasets}. In
  \bibinfo{booktitle}{\emph{2014 Data Compression Conference}}. IEEE,
  \bibinfo{pages}{203--212}.
\newblock


\bibitem[Xiao and Nagamochi(2017)]%
        {xiao2017exact}
\bibfield{author}{\bibinfo{person}{Mingyu Xiao} {and} \bibinfo{person}{Hiroshi
  Nagamochi}.} \bibinfo{year}{2017}\natexlab{}.
\newblock \showarticletitle{Exact algorithms for maximum independent set}.
\newblock \bibinfo{journal}{\emph{Information and Computation}}
  \bibinfo{volume}{255} (\bibinfo{year}{2017}), \bibinfo{pages}{126--146}.
\newblock


\bibitem[Xu et~al\mbox{.}(2009)]%
        {xu2009design}
\bibfield{author}{\bibinfo{person}{Qikai Xu}, \bibinfo{person}{Michael~R
  Schlabach}, \bibinfo{person}{Gregory~J Hannon}, {and}
  \bibinfo{person}{Stephen~J Elledge}.} \bibinfo{year}{2009}\natexlab{}.
\newblock \showarticletitle{Design of 240,000 orthogonal 25mer DNA barcode
  probes}.
\newblock \bibinfo{journal}{\emph{Proceedings of the National Academy of
  Sciences}} \bibinfo{volume}{106}, \bibinfo{number}{7} (\bibinfo{year}{2009}),
  \bibinfo{pages}{2289--2294}.
\newblock


\bibitem[Yamamoto et~al\mbox{.}(2008)]%
        {yamamoto2008large}
\bibfield{author}{\bibinfo{person}{Masahito Yamamoto}, \bibinfo{person}{Satoshi
  Kashiwamura}, \bibinfo{person}{Azuma Ohuchi}, {and} \bibinfo{person}{Masashi
  Furukawa}.} \bibinfo{year}{2008}\natexlab{}.
\newblock \showarticletitle{Large-scale DNA memory based on the nested PCR}.
\newblock \bibinfo{journal}{\emph{Natural Computing}} \bibinfo{volume}{7},
  \bibinfo{number}{3} (\bibinfo{year}{2008}), \bibinfo{pages}{335--346}.
\newblock


\bibitem[Yazdi et~al\mbox{.}(2017)]%
        {yazdi2017portable}
\bibfield{author}{\bibinfo{person}{SM Yazdi}, \bibinfo{person}{Ryan Gabrys},
  {and} \bibinfo{person}{Olgica Milenkovic}.} \bibinfo{year}{2017}\natexlab{}.
\newblock \showarticletitle{Portable and error-free DNA-based data storage}.
\newblock \bibinfo{journal}{\emph{Scientific reports}} \bibinfo{volume}{7},
  \bibinfo{number}{1} (\bibinfo{year}{2017}), \bibinfo{pages}{1--6}.
\newblock


\bibitem[Yazdi et~al\mbox{.}(2015a)]%
        {yazdi2015dna}
\bibfield{author}{\bibinfo{person}{SM~Hossein~Tabatabaei Yazdi},
  \bibinfo{person}{Han~Mao Kiah}, \bibinfo{person}{Eva Garcia-Ruiz},
  \bibinfo{person}{Jian Ma}, \bibinfo{person}{Huimin Zhao}, {and}
  \bibinfo{person}{Olgica Milenkovic}.} \bibinfo{year}{2015}\natexlab{a}.
\newblock \showarticletitle{DNA-based storage: Trends and methods}.
\newblock \bibinfo{journal}{\emph{IEEE Transactions on Molecular, Biological
  and Multi-Scale Communications}} \bibinfo{volume}{1}, \bibinfo{number}{3}
  (\bibinfo{year}{2015}), \bibinfo{pages}{230--248}.
\newblock


\bibitem[Yazdi et~al\mbox{.}(2015b)]%
        {yazdi2015rewritable}
\bibfield{author}{\bibinfo{person}{SM~Hossein~Tabatabaei Yazdi},
  \bibinfo{person}{Yongbo Yuan}, \bibinfo{person}{Jian Ma},
  \bibinfo{person}{Huimin Zhao}, {and} \bibinfo{person}{Olgica Milenkovic}.}
  \bibinfo{year}{2015}\natexlab{b}.
\newblock \showarticletitle{A rewritable, random-access DNA-based storage
  system}.
\newblock \bibinfo{journal}{\emph{Scientific reports}}  \bibinfo{volume}{5}
  (\bibinfo{year}{2015}), \bibinfo{pages}{14138}.
\newblock


\bibitem[Yin et~al\mbox{.}(2021)]%
        {Yin2021DesignOC}
\bibfield{author}{\bibinfo{person}{Qiang Yin}, \bibinfo{person}{Yanfen Zheng},
  \bibinfo{person}{Bin Wang}, {and} \bibinfo{person}{Qiang Zhang}.}
  \bibinfo{year}{2021}\natexlab{}.
\newblock \showarticletitle{Design of Constraint Coding Sets for Archive DNA
  Storage.}
\newblock \bibinfo{journal}{\emph{IEEE/ACM transactions on computational
  biology and bioinformatics}}  \bibinfo{volume}{PP} (\bibinfo{year}{2021}).
\newblock


\bibitem[Zadeh et~al\mbox{.}(2011)]%
        {zadeh2011nupack}
\bibfield{author}{\bibinfo{person}{Joseph~N Zadeh}, \bibinfo{person}{Conrad~D
  Steenberg}, \bibinfo{person}{Justin~S Bois}, \bibinfo{person}{Brian~R Wolfe},
  \bibinfo{person}{Marshall~B Pierce}, \bibinfo{person}{Asif~R Khan},
  \bibinfo{person}{Robert~M Dirks}, {and} \bibinfo{person}{Niles~A Pierce}.}
  \bibinfo{year}{2011}\natexlab{}.
\newblock \showarticletitle{NUPACK: Analysis and design of nucleic acid
  systems}.
\newblock \bibinfo{journal}{\emph{Journal of computational chemistry}}
  \bibinfo{volume}{32}, \bibinfo{number}{1} (\bibinfo{year}{2011}),
  \bibinfo{pages}{170--173}.
\newblock


\bibitem[Zhirnov et~al\mbox{.}(2016)]%
        {zhirnov2016nucleic}
\bibfield{author}{\bibinfo{person}{Victor Zhirnov}, \bibinfo{person}{Reza~M
  Zadegan}, \bibinfo{person}{Gurtej~S Sandhu}, \bibinfo{person}{George~M
  Church}, {and} \bibinfo{person}{William~L Hughes}.}
  \bibinfo{year}{2016}\natexlab{}.
\newblock \showarticletitle{Nucleic acid memory}.
\newblock \bibinfo{journal}{\emph{Nature materials}} \bibinfo{volume}{15},
  \bibinfo{number}{4} (\bibinfo{year}{2016}), \bibinfo{pages}{366--370}.
\newblock


\bibitem[Zhu et~al\mbox{.}(2008)]%
        {zhu2008avoiding}
\bibfield{author}{\bibinfo{person}{Benjamin Zhu}, \bibinfo{person}{Kai Li},
  {and} \bibinfo{person}{R~Hugo Patterson}.} \bibinfo{year}{2008}\natexlab{}.
\newblock \showarticletitle{Avoiding the Disk Bottleneck in the Data Domain
  Deduplication File System.}. In \bibinfo{booktitle}{\emph{Fast}},
  Vol.~\bibinfo{volume}{8}. \bibinfo{pages}{1--14}.
\newblock


\end{thebibliography}

\end{document}